\documentclass[useAMS, usenatbib, amssymb]{mnras}
\usepackage{psfig}
\usepackage{graphicx}
\usepackage{epsf}
\usepackage{bm}
\usepackage{lscape}
\usepackage[usenames, dvips]{color}
\title [Empirical Bolometric Correction Coefficients]
{Empirical Bolometric Correction Coefficients for Nearby Main-Sequence Stars in {\it Gaia} Era}
\author[Eker~et~al.]
       {Z. Eker $^{1}$\thanks{E-mail: eker@akdeniz.edu.tr},
F. Soydugan$^{2,3}$, S. Bilir$^{4}$, V. Bak{\i}\c{s}$^{1}$, F. Ali\c{c}avu\c{s}$^{2,3}$, S. \"Ozer$^{2,3}$,
\newauthor
 G. Aslan$^{1}$, M. Alpsoy$^{1}$, Y. K\"ose$^{1}$
\\
    $^1$Akdeniz University, Faculty of Sciences, Department of Space Sciences and 
Technologies, 07058, Antalya, Turkey\\
    $^2$Department of Physics, Faculty of Arts and Sciences, \c{C}anakkale Onsekiz 
Mart University, 17100 \c{C}anakkale, Turkey\\
    $^3$Astrophysics Research Center and Ulup{\i}nar Observatory, \c{C}anakkale 
Onsekiz Mart University, 17100, \c{C}anakkale, Turkey\\
    $^4$Istanbul University, Faculty of Science, Department of Astronomy and Space
Sciences, 34119, Istanbul, Turkey\\
}
\date{}

\pagerange{\pageref{firstpage}--\pageref{lastpage}} \pubyear{}

\begin{document}

\maketitle

\label{firstpage} 
\begin{abstract}
Nearby detached double-lined eclipsing binaries with most accurate data were studied and 290 systems were found with at least one main-sequence component having a metallicity $0.008\leq Z\leq 0.040$. Stellar parameters, light ratios, {\it Gaia} DR2 trigonometric parallaxes, extinctions and/or reddening were investigated and only 206 systems were selected eligible to calculate empirical bolometric corrections. NASA-IPAC Galactic dust maps were main source of extinctions. Unreliable extinctions at low Galactic latitudes $|b|\leq5^o$ were replaced with individual determinations, if they exist in the literature, else associated systems are discarded. Main-sequence stars of remaining systems were used to calculate bolometric corrections ($BC$) and to calibrate $BC-T_{eff}$ relation, which is valid in the range 3100-36000 K. De-reddened $(B-V)_0$ colours, on the other hand, allowed us to calibrate two intrinsic colour effective temperature relations, where the linear one is valid for $T_{eff}>10000$ K, while the quadratic relation is valid for $T_{eff}<10000$ K, that is, both are valid in the same temperature range $BC-T_{eff}$ relation is valid. New $BC$ computed from $T_{eff}$ and other astrophysical parameters are tabulated, as well.
\end{abstract}

\begin{keywords}
Stars: fundamental parameters -- Stars: binaries: eclipsing -- Stars: binaries: spectroscopic 
-- Sun: general; Astrophysics - Solar and Stellar Astrophysics
\end{keywords}

\section{Introduction}
Outstanding improvements in stellar astrophysics have been recorded in recent years. High sensitivity images taken with satellite telescopes, data obtained with advanced new technology telescopes and high-resolution spectrographs have an important share in this. While advanced analysis techniques were added to these advances, the basic astrophysical parameters began to be obtained very precisely (especially close to 1\% precision in the mass and radius of double-lined detached binaries). On the other hand, it is important for many purposes to obtain the total luminosity or bolometric brightness of stars. Bolometric corrections ($BC$) are necessary to derive the total luminosity. Luminosity, however, is typically measured in a given bandpass (or a few bandpasses). To pass from luminosity in a given band to total luminosity, $BC$s are needed. 

Historically, a $BC$ for a star is defined as the difference between its bolometric and visual magnitudes as, 

\begin{equation}
BC = M_{Bol}-M_{V} = m_{Bol}-m_{V},
\end{equation}
where subscripts ``Bol'' and ``V'' imply bolometric and visual bandwidths if $M$ is the absolute and $m$ is the apparent magnitudes. Here, the preference of the visual band is the convention because throughout the history visual magnitudes are mostly used and mostly available.
 
Since there is no telescope or a detector to observe a star at all wavelengths and because the bolometric magnitude of the star is related to its luminosity by the relation,   

\begin{equation}
M_{Bol}=M_{Bol,\odot} - 2.5\log L/L_{\odot},
\end{equation}
where $M_{Bol,\odot}$ and $L_{\odot}$ are the reference quantities representing the bolometric magnitude and the luminosity of the Sun, then $BC$ is a correction for the missing luminosity. Similarly, apparent magnitude of the star is related to its flux ($f$) just above the atmosphere by the relation

\begin{equation}
m_{V} = -2.5\log f_V + C_V,
\end{equation}
where $C_V$ and $f_V$ are the zero point constant and the flux received from this star, thus $BC$ may also be understood as the missing flux which cannot be observed due to limited observable bandwidth. 

There is a large body of works on this topic. Notable early ones are summarized by \citet{Kuiper38} and \citet{Popper59}. Most works used Spectral Energy Distribution (SED) in different bands, typically given in tabular forms \citep*{McDonald52, Heintze73, Code76, Malagnini85, Cayrel97, Bessell98, Girardi08}, and some used radiometric measurements \citep{Pettit28, Wildey63}, while some preferred classical photometric methods to derive $BC$s together with intrinsic colours \citep{Johnson64, Johnson66, Flower77, Flower96, Sung13}. Recently, synthetic stellar photometry and theoretical spectral libraries has been also preferred to drive various kinds of $BC$s \citep{Casagrande18, Chen19}.    

In recent years, one of the most used paper is that of \citet{Flower96}, who gives three empirical $BC-T_{eff}$ relations each valid at three different temperature ranges, but usable for all luminosity classes \citep[see also,][]{Torres10}. Here we want to improve upon the work of \citet{Flower96} including recently compiled more reliable data  in order to test the claim of if a single empirical $BC-T_{eff}$ relation would be valid or not for all stellar temperatures and for all spectral classes. We take the advantage of {\it Gaia}, which provides the most precise and accurate stellar parallaxes ever measured, and the numerous more accurate data which was gathered since then. For this study, we start with obtaining empirical $BC$ coefficients of nearby main-sequence stars.

Using the most reliable stellar basic parameters ($M$, $R$, $T_{eff}$), and {\it Gaia} DR2 trigonometric parallaxes of the detached double-lined eclipsing binaries \citep{Gaia16, Gaia18} and interstellar extinction ($A_V$) partly from the NASA/IPAC Galactic dust maps \citep{Schlafly11} partly from private determinations, $BC$ coefficients for Solar neighborhood main-sequence stars are determined and a new $BC-T_{eff}$ relation is calibrated in this study. This new relation is more practical (with less number of digits) compared to the coefficients of the rectified relations by \citet{Torres10} and it is a single fourth degree polynomial valid for the temperature range 3100-36000 K. 

\section{Data}

An absolute bolometric magnitude of a star could be computed by Eq. (2) using the reference quantities $M_{Bol}=4.74$ mag and $L_{\odot} =3.828\times10^{26}$ W according to IAU 2015 General Assembly resolution B2\footnote{https://www.iau.org/static/resolutions/IAU2015\_English.pdf}. However, a luminosity of a star is not one of its directly observable parameters, but it is an empirical parameter which has to be computed from observable parameters: the radius ($R$) and the effective temperature ($T_{eff}$) using the following relation 

\begin{equation}
{\frac{L}{L_{\odot}}}=\left({\frac{R}{R_{\odot}}}\right)^2\left({\frac{T_{eff}}{T_{eff,\odot}}}\right)^4,
\end{equation}
in accord with the Stefan-Boltzmann law. Since stellar temperatures are in Kelvin and stellar radii are in Solar units, $T_{eff,\odot}=5772$ K is also needed according to IAU 2015 General Assembly resolution B3.

It is clear that when computing an empirical $BC$ of a star, a reliable observational $R$ and a trustable $T_{eff}$ are needed in the first step to obtain its $M_{Bol}$. Next a reliable apparent visual magnitude ($m_V$), a distance ($d$) and an extinction ($A_V$) in the visual band are necessary for the second step according to

\begin{equation}
m_V - M_V = 5\log d - 5 + A_V,
\end{equation}
from where visual absolute magnitude ($M_V$) of the star could be extracted. Finally $M_{Bol}$ and $M_V$ would be ready to apply Eq. (1).

\subsection{Selecting stars for the first step}

Since most accurate stellar parameters ($M$, $R$, $T_{eff}$) come from simultaneous solutions of observed light and radial velocity curves of detached double-lined eclipsing binaries, main source of data for this study is ``The Catalogue of Stellar Parameters from the Detached Double-lined Eclipsing Binaries in the Milky Way'' which was originally compiled by \citet{Eker14}. Although the steps described are applicable to all stars, only main-sequence stars were chosen for this study because the number of stars out of  the main-sequence in the catalogue are insufficient to explore their $BC-T_{eff}$ relations.

The original catalogue of \citet{Eker14} is updated and the number of systems is increased to 319 (318 SB2 and one SB3) by \citet{Eker18}, who selected 509 stars with $M$ and $R$ both accurate up to 15\% and metallicities ($0.008\leq Z\leq 0.040$) in order to calibrate interrelated main-sequence mass-luminosity (MLR), mass-radius (MRR) and mass-effective temperature (MTR) relations for the Galactic nearby stars. Therefore, the list already prepared by \citet{Eker18} is considered to be the initial list for the present study.

Unfortunately, not all stars with $M$ and $R$ accuracies within 15\% and metallicities within $0.008\leq Z\leq 0.040$ in the catalogue have published $T_{eff}$. This is because some authors, such \citet{Young06}, \citet{Shkolnik08}, \citet{Helminiak09} and \citet{Sandquist13}, prefer to give temperature ratios rather than individual temperatures of the components in their publication. 53 stars (25 SB2 and 1 SB3) without published $T_{eff}$ were eliminated right at the beginning; that is, not included in the calibrations of interrelated MLR, MRR and MTR. For this study, we have re-explored those 53 stars and found 34 of them fulfilling the both conditions that $M$ and $R$ are both accurate up to 15\% and metallicities $0.008\leq Z\leq 0.040$ on the main-sequence region of the $\log M-\log R$ diagram defined by zero age main-sequence (ZAMS) and terminal age main-sequence (TAMS) lines according to PARSEC stellar evolution models \citep{Bressan12}. Now, these stars too are included in the list of stars collected for the first step. This is because; it is now possible to assign them a reliable $T_{eff}$ according to the basic relations MLR, MRR and MTR. 

Recently, \citet{Graczyk19} studied the global zero-point shift between the photometric parallaxes of 81 detached eclipsing binaries and {\it Gaia} Data Release 2 (DR2) trigonometric parallaxes \citep{Gaia18}. 72 of those binaries are already in our catalog. The paper of \citet{Graczyk19} contained more, that is, additional seven systems (AL Ari, AL Dor, V530 Ori, KX Cnc, V923 Sco, TYC 7091-888-1, EPIC 211409263) with accurately determined $R$ and $T_{eff}$ from light and radial velocity curves using Wilson-Devinney code version 2007 \citep{vanHamme07}. We decided also to include those seven systems into the list of stars for the first step. Finally, the number of stars is maximized to 550 stars, which their absolute bolometric magnitude could be computed from their $R$ and $T_{eff}$ according to (2) and (4) for the first step of computing $BC$.

\subsection{Selecting systems for the second step}
The number of stars is maximized in the first step above because this number would be decreased greatly in the second step. The loss is inevitable because it is not possible to calculate absolute visual magnitude $M_V$ of a star if any of the observational parameters $m_V$, $d$ or $A_V$ is absent. Moreover, rather than counting individual stars, counting binary systems is more meaningful in the second step. This is because if an extinction ($A_V$) or a distance ($d$) is missing, it is missing for the system, not only for one star. Furthermore, apparent visual magnitude of a component ($m_V$) is available only if a reliable light ratio ($L_2/L_1$) of the components in the visual band is available. Thus, systems without ($L_2/L_1$) in the $V$ band are also lost. Therefore, chosen stars organized as systems. Consequently, we have counted 290 systems, which have at least one component on the main-sequence. 

We went through each system and have been able to select 206 systems with complete data to calculate both $M_{Bol}$ and $M_V$, that is, $BC$ is computable. The basic observational parameters of 206 systems are given in Table 1. The columns are self-explanatory and organized as: ID number, name of the system, equatorial coordinates ($\alpha$, $\delta$; International Celestial Reference System in J2000.0), Galactic coordinates ($l$, $b$), spectral types, orbital periods ($P$), references, $B$ magnitudes and errors, $V$ magnitudes and errors, references, interstellar extinction in $V$ band ($A_V$) and its error, references, trigonometric parallaxes and errors, references.

NASA/IPAC Galactic Dust Reddening and Extinction maps\footnote{https://irsa.ipac.caltech.edu/applications/DUST/} are our main source for $A_V$ data listed in Table 1. This web site  provides an extinction map, the corresponding 100 micron intensity, and dust temperature, along with statistics for each if Galactic coordinates of an object is provided. A $E(B-V)$ colour-excess  and associated error towards the direction of a star, from the Sun up to the Galactic border, due to the dust in the Milky Way disk are taken from \citet{Schlafly11}, who assume a single standard extinction law $A_V=3.1 E(B-V)$. Consequently, corresponding  total extinction (from Sun up to Galactic border) and associated error for the $V$ band are known. The total extinction ($A_{\infty}(b)$) is then reduced for the distance of the star by using the relation of \citet{Bahcall80}, 

\begin{equation}
A_{d}(b)=A_{\infty}(b)\Biggl[1-\exp\Biggl(\frac{-\mid d~\sin(b)\mid}{H}\Biggr)\Biggr],
\end{equation}  
where $b$ is the Galactic latitude, $d$ is the distance of the system from the Sun. $H$ is the Galactic dust scaleheight \citep[$H$=125 pc;][]{Marshall06}, Only, the two systems, $\beta$ Aur and V2083 Cyg, do not have {\it Gaia} trigonometric parallax measurements because they are too bright for {\it Gaia} detectors. We did not want to discard them, instead use them with their {\it Hipparcos} trigonometric parallaxes \citep{vanLeeuwen07}. There are 12 more systems which we also preferred to use their {\it Hipparcos} trigonometric parallaxes. Those are the systems which are brighter than $V=8$ mag and having hot components (spectral types A or earlier).  

It is clear by Eq. (6) that, extinctions of the systems with low-Galactic latitudes ($|b|<5^o$) are unreliable, especially if $d$ is large ($d>100$ pc). Such systems are more likely to be embedded in the dust concentrated in the Galactic plane. Therefore, for the systems with relatively low-Galactic latitudes and for the systems with large reddening, we have searched in the literature if their $E(B-V)$ colour excess is measured by other authors. If there is an observationally determined value of $A_V$ or $E(B-V)$, we have preferred them rather than our estimates according to NASA/IPAC Galactic Dust Reddening and Extinction maps. Some systems with hotter components, spectral types A or earlier, we could not find any privately determined $A_V$ or $E(B-V)$, but we were able to use $Q$ method \citep{Johnson53} to estimate their $E(B-V)$ and $A_V$. If $Q$ method is not applicable (systems of late spectral types) and if there is no published $A_V$ or $E(B-V)$ for a system, then it is discarded because of unreliable $A_V$. Among the 206 system listed in Table 1, there are 131 systems with $A_V$, which is estimated from the Galactic dust maps, and rest (75 systems) are from the $E(B-V)$ colour-excess measurements by the authors referenced in the table. Extinctions relying on $Q$ method are referenced to ``this study''. 

Table 2 gives the rest of the other parameters which are needed in computing $BC$ values. The columns of Table 2 are organized as: ID number, name of the system, primary mass and its error, secondary mass and its error, primary radius and its error, secondary radius and its error, references, primary effective temperature and its error, secondary effective temperature and its error, references, light ratio ($L_2/L_1$) in $B$ band, $B$ contribution of the third body if exists, light ratio ($L_2/L_1$) in $V$ band, $V$ contribution of the third body if exists, references.

\begin{landscape}
\begin{table}
\setlength{\tabcolsep}{1pt}
{\tiny
  \centering
  \caption{Basic observational parameters of the detached double-lined eclipsing systems selected.}
    \begin{tabular}{clcccclllcclclcl}
\hline
 ID & Name          & $\alpha$ & $\delta$        & $l$    & $b$    &\multicolumn{1}{c}{Spt Type}  &\multicolumn{1}{c}{$P$}      &\multicolumn{1}{c}{Reference}            & $B$              & $V$              & \multicolumn{1}{c}{Reference}            & $A_V$           & \multicolumn{1}{c}{Reference}           & $\varpi$             & \multicolumn{1}{c}{Reference}\\
    &       & (hh:mm:ss) & (dd:mm:ss) & ($^o$) & ($^o$) &      & \multicolumn{1}{c}{(day)}    &  &         (mag)      &  (mag) &   & (mag) &  & (mas) &  \\
\hline
  1 & V421 Peg      & 00:07:02.00 & +22:50:40.03 & 109.75 & -38.89 & F1V + F2V     & 3.087566    & 2007A\&A...474..653V &         ---      &  8.280$\pm$0.010 & 2016NewA...46...47O  & 0.158$\pm$0.005 & 2011ApJ...737..103S & 6.3806$\pm$0.0543 & 2018A\&A...616A...1G  \\
  2 & DV Psc        & 00:13:09.20 & +05:35:43.06 & 105.72 & -55.99 & K4V + M1V     & 0.30853609  & 2007MNRAS.382.1133Z  & 11.604$\pm$0.010 & 10.621$\pm$0.010 & 2000A\&A...355L..27H & 0.012$\pm$0.001 & 2011ApJ...737..103S & 0.0120$\pm$0.0787 & 2018A\&A...616A...1G  \\
  3 & MU Cas        & 00:15:51.56 & +60:25:53.64 & 118.55 &  -2.14 & B5V + B5V     & 9.652929    & 2004AJ....128.1840L  & 11.112$\pm$0.009 & 10.808$\pm$0.007 & 2019ApJ...872...85G  & 1.249$\pm$0.027 & 2011ApJ...737..103S & 1.2490$\pm$0.0397 & 2018A\&A...616A...1G  \\
  4 & GSC 4019-3345 & 00:22:45.37 & +62:20:05.50 & 119.61 &  -0.35 & A4 V + A4 V   & 4.077304    & 2013PASA...30...26B  & 12.550$\pm$0.009 & 12.150$\pm$0.008 & 2013PASA...30...26B  & 0.880$\pm$0.100 & This study          & 0.9513$\pm$0.0316 & 2018A\&A...616A...1G  \\
  5 & YZ Cas        & 00:45:39.08 & +74:59:17.06 & 122.55 &  12.12 & A2m + F2V     & 4.4672235   & 2014MNRAS.438..590P  &  5.715$\pm$0.026 &  5.660$\pm$0.015 & 2019ApJ...872...85G  & 0.195$\pm$0.008 & 2011ApJ...737..103S & 0.1950$\pm$0.0939 & 2018A\&A...616A...1G  \\
... & ...           & ...         & ...          & ...    &    ... & ...           & ...         & ...                  & ...              &    ...           &           ...        & ...             &  ...                & ...               &  ...                  \\
... & ...           & ...         & ...          & ...    &    ... & ...           & ...         & ...                  & ...              &    ...           &           ...        & ...             &  ...                & ...               &  ...                  \\
... & ...           & ...         & ...          & ...    &    ... & ...           & ...         & ...                  & ...              &    ...           &             ...      & ...             &  ...                & ...               &  ...                  \\
200 & AR Cas        & 23:30:01.94 & +58:32:56.11 & 112.47 &  -2.66 & B4V + A6V     & 6.066317    & 1999A\&A...345..855H &  4.777$\pm$0.004 &  4.893$\pm$0.004 & 2003ARep...47..551K  & 0.350$\pm$0.030 & 2003ARep...47..551K & 0.3500$\pm$0.2047 & 2018A\&A...616A...1G  \\
201 & V731 Cep      & 23:37:43.55 & +64:18:11.20 & 115.05 &   2.57 & B8.5V + A1.5V & 6.068456    & 2008MNRAS.390..399B  & 10.630$\pm$0.008 & 10.540$\pm$0.008 & 2008MNRAS.390..399B  & 0.403$\pm$0.093 & This study          & 0.4030$\pm$0.0305 & 2018A\&A...616A...1G  \\
202 & IT Cas        & 23:42:01.40 & +51:44:36.80 & 112.13 &  -9.67 & F6V + F6V     & 3.8966721   & 1997AJ....114.1206L  & 11.640$\pm$0.019 & 11.150$\pm$0.039 & 1997AJ....114.1206L  & 0.180$\pm$0.030 & 2016AJ....152..180S & 2.0149$\pm$0.0358 & 2018A\&A...616A...1G  \\
203 & BK Peg        & 23:47:08.46 & +26:33:59.92 & 105.53 & -34.12 & F8            & 5.48991046  & 1983AJ.....88.1242P  & 10.540$\pm$0.041 &  9.982$\pm$0.010 & 2019ApJ...872...85G  & 0.152$\pm$0.004 & 2011ApJ...737..103S & 0.1520$\pm$0.0526 & 2018A\&A...616A...1G  \\
204 & AP And        & 23:49:30.71 & +45:47:21.25 & 111.78 & -15.74 & F6 + F8       & 1.587291156 & 1984ApJS...54..421G  & 11.606$\pm$0.057 & 11.074$\pm$0.085 & 2015AAS...22533616H  & 0.270$\pm$0.020 & 2016AJ....152..180S & 0.2700$\pm$0.0415 & 2018A\&A...616A...1G  \\
205 & AL Scl        & 23:55:16.58 & -31:55:17.28 &   8.14 & -76.89 & B6V + B9V     & 2.445083    & 1987A\&A...179..141H &  5.985$\pm$0.014 &  6.070$\pm$0.009 & 2000A\&A...355L..27H & 0.100$\pm$0.010 & This study          & 4.2680$\pm$0.1273 & 2018A\&A...616A...1G  \\
206 & V821 Cas      & 23:58:49.17 & +53:40:19.82 & 115.10 & -8.4 & A1.5V + A4V     & 1.7697999   & 2009MNRAS.395.1649C  &  8.402$\pm$0.029 &  8.286$\pm$0.017 & 2019ApJ...872...85G  & 0.217$\pm$0.009 & 2011ApJ...737..103S & 0.2170$\pm$0.0376 & 2018A\&A...616A...1G  \\
\hline
    \end{tabular}
}
\end{table}

\begin{table}
\setlength{\tabcolsep}{3pt}
{\tiny
  \centering
  \caption{Physical parameters of selected systems which are necessary in computing their $BC$.}
    \begin{tabular}{clcccclcclclccl}
\hline
   &                &                &                           &                 &                 &                     &               &               &                     & \multicolumn{2}{c}{$B$ Band} & \multicolumn{2}{c}{$V$ Band} &           \\
ID & Name           & $M_1$          &          $M_2$            & $R_1$           & $R_2$           & \multicolumn{1}{c}{Reference}           & $T_1$         & $T_2$         & \multicolumn{1}{c}{Reference}           & $L_2/L_1$ & $L_3$    & $L_2/L_1$ & $L_3$    & Reference \\
   &                & ($M_{\odot}$)  &       ($M_{\odot}$)       & ($R_{\odot}$)   & ($R_{\odot}$)   &                     & (K)           & (K)           &                     &           &          &           &          &           \\
\hline
  1 & V421 Peg      & 1.594$\pm$0.029 & 1.356$\pm$0.029 & 1.584$\pm$0.028 & 1.328$\pm$0.029 & 2016NewA...46...47O & 7250$\pm$80   & 6980$\pm$120  & 2016NewA...46...47O & ---    & --- & 0.6023 & --- & 2016NewA...46...47O \\
  2 & DV Psc        & 0.677$\pm$0.019 & 0.475$\pm$0.010 & 0.685$\pm$0.030 & 0.514$\pm$0.020 & 2014PASA...31...24E & 4450$\pm$8    & 3614$\pm$8    & 2007MNRAS.382.1133Z & 0.0893 & --- & 0.1254 & --- & 2014AJ....147...50P \\
  3 & MU Cas        & 4.657$\pm$0.100 & 4.575$\pm$0.090 & 4.192$\pm$0.050 & 3.671$\pm$0.040 & 2014PASA...31...24E & 14750$\pm$500 & 15100$\pm$500 & 2004AJ....128.1840L & 0.7982 & --- & 0.7957 & --- & 2019ApJ...872...85G  \\
  4 & GSC 4019-3345 & 1.920$\pm$0.010 & 1.920$\pm$0.010 & 1.760$\pm$0.050 & 1.760$\pm$0.050 & 2013PASA...30...26B & 8600$\pm$310  & 8600$\pm$570  & 2013PASA...30...26B & 1.0121 & --- & 1.0161 & --- & 2013PASA...30...26B \\
  5 & YZ Cas        & 2.263$\pm$0.012 & 1.325$\pm$0.007 & 2.525$\pm$0.011 & 1.331$\pm$0.006 & 2014MNRAS.438..590P & 9520$\pm$120  & 6880$\pm$240  & 2014MNRAS.438..590P & 0.0610 & --- & 0.0880 & --- & 2019ApJ...872...85G  \\
... & ...           &  ...            &  ...            &  ...            &   ...           &  ...                & ...           & ...           & ...                 & ...    & ... & ...  & ... &  ...         \\
... & ...           &  ...            &  ...            &  ...            &   ...           &  ...                & ...           & ...           & ...                 & ...    & ... & ...  & ... &  ...         \\
... & ...           &  ...            &  ...            &  ...            &   ...           &  ...                & ...           & ...           & ...                 & ...    & ... & ...  & ... &  ...         \\
200 & AR Cas        & 5.900$\pm$0.200 & 1.860$\pm$0.060 & 4.860$\pm$0.100 & 1.590$\pm$0.030 & 2014PASA...31...24E & 16800$\pm$200 & 8250$\pm$100  & 2003ARep...47..551K & 0.0205 & 1.14& 0.0304 & 1.71& 2003ARep...47..551K \\
201 & V731 Cep      & 2.577$\pm$0.098 & 2.017$\pm$0.084 & 1.823$\pm$0.030 & 1.717$\pm$0.025 & 2014PASA...31...24E & 10700$\pm$200 & 9265$\pm$220  & 2008MNRAS.390..399B & 0.5649 & --- & 0.6129 & --- & 2008MNRAS.390..399B \\
202 & IT Cas        & 1.330$\pm$0.009 & 1.328$\pm$0.008 & 1.603$\pm$0.015 & 1.569$\pm$0.040 & 2014PASA...31...24E & 6470$\pm$110  & 6470$\pm$110  & 1997AJ....114.1206L & 0.9685 & --- & 0.9646 & --- & 1997AJ....114.1206L \\
203 & BK Peg        & 1.414$\pm$0.007 & 1.257$\pm$0.005 & 1.985$\pm$0.008 & 1.472$\pm$0.017 & 2014PASA...31...24E & 6265$\pm$85   & 6320$\pm$30   & 2010A\&A...516A..42C& 0.5780 & --- & 0.5721 & --- & 2019ApJ...872...85G  \\
204 & AP And        & 1.277$\pm$0.004 & 1.251$\pm$0.004 & 1.234$\pm$0.006 & 1.195$\pm$0.005 & 2014AJ....147..148L & 6565$\pm$150  & 6495$\pm$150  & 2014AJ....147..148L & 0.8857 & --- & 0.8904 & --- & 2014MNRAS.437.3718Z \\
205 & AL Scl        & 3.617$\pm$0.110 & 1.703$\pm$0.040 & 3.241$\pm$0.050 & 1.401$\pm$0.020 & 2014PASA...31...24E & 13550$\pm$350 & 10300$\pm$360 & 1987A\&A...179..141H& 0.0417 & --- & 0.0526 & --- & 1981Ap\&SS..74...83G \\
206 & V821 Cas      & 2.025$\pm$0.066 & 1.620$\pm$0.058 & 2.308$\pm$0.028 & 1.390$\pm$0.022 & 2014PASA...31...24E & 9400$\pm$400  & 8600$\pm$400  & 2009MNRAS.395.1649C & 0.2550 & --- & 0.2830 & --- & 2019ApJ...872...85G  \\
\hline
    \end{tabular}
}
\end{table}
\end{landscape}

Mass and radius collected from older references are homogenized and re-evaluated using recently updated and more accurate constants $GM_{\odot}=1.3271244\times10^{20}$ m$^3$s$^{-2}$ \citep{Standish95} and $R_{\odot}=6.9566\times10^8$m \citep{Haberreiter08} by \citet{Eker14}. Therefore, absolute parameters $M$, $R$ coming from the older references are given a single reference. Interested readers may follow the references given in \citet{Eker14} for the original published values. The $GM_{\odot}$ value adopted by \citet{Eker14, Eker15} is the same as the $GM_{\odot}$ value adopted by IAU 2015 General Assembly resolution B3. However, the Solar radius ($R_{\odot}$) is 0.057\% smaller than $R_{\odot}=6.957\times10^{8}$ m which is adopted by IAU 2015. Such a small difference is definitely negligible besides the uncertainties of the observed stellar radii.

The basic astrophysical parameters ($M$, $R$ and $T_{eff}$) of the secondary components of TZ for, Z Her, HD147827, BD-20 5728, EPIC 211409263, NSVS 11868841, NSVS 06502726, NSVS 02502726 and primary components of TYC 7091-888-1, TYC 176-2950-1, V432 Aur and V380 Cyg are not listed in Table 2 because they are not main-sequence stars, thus, they are discarded from this study. 

The most critical requirement for a system to be included in this study is to have light ratio ($L_2/L_1$) in the $V$ band. Without it, apparent brightness' of the components in the $V$ band cannot be computed, so their $BC$ coefficients. Therefore, all 206 systems have a light ratio ($L_2/L_1$) for the $V$ band. Originally, we were also interested in $L_2/L_1$ for the $B$ band in order to estimate un-reddened $B-V$ colours of the components. However, $L_2/L_1$ for the $B$ band cannot be found for 60 systems. Therefore, only the values of existing $L_2/L_1$ for the $B$ band are listed in Table 2. Correspondingly, the systemic brightness' for $B$ band for these systems are also left empty for the same reason in Table 1, not because we could not found $B$ magnitudes for these systems in the literature.  

\subsection{Main-sequence confirmation of the sample}

The positions of 400 stars (194 binaries, 8 primaries, and 4 secondaries) are shown on a $\log M - \log R$ diagram in Fig. 1a, where the ZAMS and TAMS lines of PARSEC stellar evolution models \citep{Bressan12} marks the region of the main-sequence stars with metallicities $0.008\leq Z\leq 0.040$. So, Fig. 1a confirms that the components (400 stars) of the binaries listed in Table 1 and Table 2 are main-sequence stars having metallicities within the limits $0.008\leq Z\leq 0.040$. All must be within the Galactic disc in the Solar neighborhood according to their Galactic coordinates and trigonometric parallaxes listed. The positions of discarded non main-sequence components (12 stars) are shown on a similar $\log M-\log R$ diagram in Fig. 1b. Fig. 1b confirms that discarded stars are not on the main-sequence.

\begin{figure}
\begin{center}
\includegraphics*[scale=0.65,angle=0]{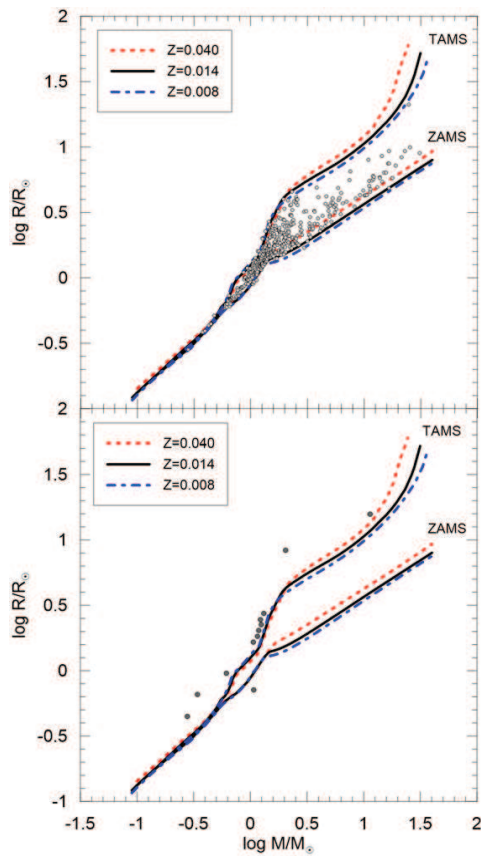}
\caption[] {a) Positions of 400 stars (194 binaries, eight primary and four secondary), b) positions of discarded components (eight secondary, four primary) on the $M-R$ diagram. ZAMS and TAMS lines of PARSEC stellar evolution models (dash-dot $Z=0.008$, solid Solar, dotted $Z$=0.040) mark the region of the main-sequence stars.}
\end{center}
\end{figure}

\subsection{Propagation of observational uncertainties to $BC$}

According to Eq. (1) there are two groups of observational uncertainties to be propagated. The first group propagates trough $M_{Bol}$ and the other group propagates trough $M_V$. Therefore, according to Eq. (1), we can write for an uncertainty of a $BC$, 

\begin{equation}
\Delta BC = \sqrt{(\Delta M_{Bol})^2+(\Delta M_V)^2}.
\end{equation} 

Eqs. (2) and (4) indicates that the first group of uncertainties is included in 

\begin{equation}
\Delta M_{Bol} = 2.5\log e \sqrt{\left(2\frac{\Delta R}{R}\right)^2+\left(4\frac{\Delta T}{T}\right)^2.}
\end{equation} 
Since radius ($R_{\odot}$) and effective temperature ($T_{eff,\odot}$) of the Sun are just constants, Eq. (8) assumes no uncertainty contribution from them. The contributions are from the relative uncertainties of radii and effective temperatures of the stars in concern. Relative uncertainty of temperature usually dominates because relative errors of temperatures are usually bigger than relative errors of radii and a temperature uncertainty contributes two times even if the relative uncertainties are the same because of the power of the temperature is twice of the power of radius as appeared in the Stefan-Boltzmann law. 

According to Eq. (5), the second group of uncertainties is included in

\begin{equation}
\Delta M_V = \sqrt{\left(\Delta m_V \right)^2+\left(5\log e\frac{\sigma_\varpi}{\varpi}\right)^2+ \left(\Delta A_V \right)^2.}
\end{equation} 
Basically, there are three kinds of observational uncertainties that determine the uncertainty of $M_V$: Uncertainty of the $V$ magnitude of the component ($\Delta m_V$), uncertainty of interstellar extinction ($\Delta A_V$) and the uncertainty of the distance ($\sigma_\varpi/\varpi$). Because, the relative uncertainty of a distance is same as the relative uncertainty of the parallax, and the relative uncertainties of trigonometric parallaxes are given in Table 1, the relative uncertainty of a trigonometric parallax is entered in Eq. (9). The coefficient $5\log e = 2.1715$ is inevitable because of Eq. (5). For simplicity, we have assumed that $V$ magnitude uncertainties of the components are the same as the $V$ magnitude uncertainty of the system. This is a reasonable assumption because considerable number of systems given light ratio ($L_2/L_1$) in this study is found without uncertainties. However, systemic brightness uncertainties are already in Table 1. Since both $\Delta m_V$ and $\Delta A_V$ are usually about $\pm 0.01$ or smaller, trigonometric parallax uncertainties dominates. Not only this but also because of $5\log e = 2.1715$ values, the weight of parallax contributions to $\Delta M_V$ are approximately doubled.  

The distribution of {\it Gaia} relative parallax errors for the systems in our list are shown in Fig. 2. It has been known that if parallax precisions worse than 10\%, the error distribution in parallaxes creates a bias in {\it Gaia} distances \citep{Bailer-Jones15, Bailer-Jones18}. Since 95\% of distances in this study have relative parallax errors less than nine per cent, the distances used in this study apparently are not affected from the bias.

\begin{figure}
\begin{center}
\includegraphics[scale=0.50,angle=0]{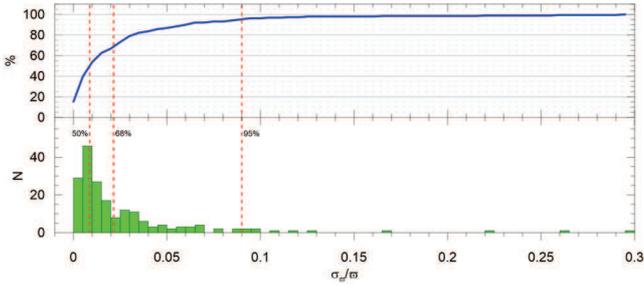}
\caption[] {Cumulative (above) and histogram (below) distributions of relative parallax errors of {\it Gaia} distances.}
\end{center}
\end{figure}

\section {Results}
\subsection{Determination of missing light ratios ($L_2/L_1$)}

The light ratio ($L_2/L_1$) is an essential parameter to obtain component apparent brightness' from the apparent brightness of the system. It can be seen in the right most columns Table 1 that there exist 47 systems, which are given references to \citet{Graczyk19} from whom light ratios in $B$ and $V$ bands are taken. Associated to 33 stars, the next most popular reference for the light ratios is ``this study'', which implies the light ratio (or ratios) is obtained by solving eclipsing light curve of the system. If these two references are absent, the number of systems would have reduced from 206 to 126. That is, statistical reliability of current results would have been greatly (almost 50\%) reduced.      

We went through original references of 550 stars collected in the first step and collected primarily the light ratio ($L_2/L_1$) in the $V$ band, and if exist in the $B$ band. We faced up the reality that not all simultaneous LC and RV solutions of detached double lined spectroscopic binaries are done in the Johnson photometry. There is great deal of systems that their solutions are done in the Str\"omgren photometry and some of the other systems like Kepler photometry, which are lost automatically because we are interested in only the light ratios in the $V$ band. Moreover, we run into publications such that the light ratio of a system is obtained but it is not given. Only physical parameters and/or other light curve solutions are given. Some of such publications do not even give brightness of the system which is also needed. 

Aiming to keep the loss of systems at a minimum level, we have requested $L_2/L_1$ from considerable number authors by private communications. V781 Per, VZ Cep, AP And \citep{Zola14}, GJ 3236 \citep{Irwin09}, V1236 Tau \citep{Bayless06}, V404 CMa \citep{Rozyczka09}, Corot 102932176 \citep*{Lazaro15}, UV Leo \citep{Kjurkchieva07}, EM Car \citep*{Cicek17}, eta Mus, GG Lup \citep*{Budding15, Budding17}, EK Cep \citep{Antonyuk09} and BW Aqr \citep{Volkov14} are the systems which we have requested $L_2/L_1$ information, if available. If not, at least $V$ and/or $B$ data are asked and got a positive reply. V1200 Cen, GU Boo, and OO Peg are the systems we did not get any reply. 

Some systems in our list contained the $V$ data printed in the paper itself or in public ASAS data archive\footnote{
http://www.astrouw.edu.pl/\~gp/asas/}. All those systems, which we could not find $L_2/L_1$, but got their data, are solved in this  study using Wilson Devinney code \citep{Wilson14} in common way \citep*[e.g.][]{Bakis18, Soydugan13}. Since the light and radial velocity curves of all these 33 systems were already analyzed in literature, the effective temperature of the primary ($T_1$) and the mass ratios of the system ($q$) were adopted from Table 2 and kept fixed during the analysis. The bolometric albedo, gravitational and limb darkening coefficients of the components were used in the same way as the values used in the analyses given in the literature. The third light contribution was also tested during the analysis. Light curves were solved in Mode 2, which corresponds to detached binaries in Wilson-Devinney code. The final parameters are given in Table 3, where the columns are self explanatory: ID number, name of the system, orbital period, temperature of primary, temperature of secondary, mass ratio, fractional radius of primary and secondary, light ratio in $V$-band, third light in $V$, light ratio in $B$-band, third light in $B$ and reference.         

\begin{landscape}
\begin{table}[htbp]
\setlength{\tabcolsep}{5pt}
{\tiny
  \centering
  \caption{Light ratios ($L_2/L_1$) from simultaneous light and radial curves analysis of 33 detached binaries that do not have light ratio information in their literature.}
    \begin{tabular}{rlccccccccccl}
\hline
   &                         &          &       &               &            &              &                   & \multicolumn{2}{c}{in $V$ filter} & \multicolumn{2}{c}{in $B$ filter}  &  \\
ID & System                  & $P$      & $T_1$ & $T_2$         & $q$        & $r_1$        & $r_2$             & $L_2/L_1$     &   $L_3$         &  $L_2/L_1$ & $L_3$ & Reference \\
   &                         &(days)    &  (K)  & (K)           & ($M_2/M_1$)& ($R_1/a$)    &  ($R_2/a$)        &    &  (\%)  &      &  (\%)  &  \\
\hline 
 1 & 2MASS J01132817-3821024 & 0.445596 &  3750 &  3112$\pm$200 & 0.727 & 0.2380$\pm$0.0002 & 0.1780$\pm$0.0002 & 0.131$\pm$0.020 & ---          & ---           & ---           & ASAS \\
 2 & V505 Per                & 4.22202  &  6512 &  6462$\pm$201 & 0.986 & 0.0861$\pm$0.0001 & 0.0847$\pm$0.0001 & 0.951$\pm$0.010 & ---          & 0.93$\pm$0.01 & ---           & \citet{Tomasella08}\\
 3 & TYC 4749-560-1          & 1.622219 &  5340 &  5125$\pm$200 & 0.993 & 0.1233$\pm$0.0005 & 0.1211$\pm$0.0005 & 0.993$\pm$0.010 & ---          & ---           & ---           & \citet{Helminiak11}\\
 4 & V1236 Tau               & 2.58791  &  4200 &  4133$\pm$250 & 0.978 & 0.0829$\pm$0.0010 & 0.0869$\pm$0.0005 & 1.039$\pm$0.010 & ---          & ---           & ---           & \citet{Bayless06} \\
 5 & V432 Aur                & 3.081745 &  6080 &  6685$\pm$85  & 0.883 & 0.2097$\pm$0.0002 & 0.1049$\pm$0.0002 & 0.372$\pm$0.020 & ---          & 0.43$\pm$0.02 & ---           & \citet{Siviero04} \\
 6 & V1388 Ori               & 2.18706  & 20500 & 18900$\pm$500 & 0.695 & 0.3400$\pm$0.0004 & 0.2283$\pm$0.0004 & 0.344$\pm$0.008 & ---          & ---           & ---           & ASAS \\
 7 & V404 CMa                & 0.45178  &  4200 &  3940$\pm$20  & 0.879 & 0.2603$\pm$0.0001 & 0.2462$\pm$0.0001 & 0.620$\pm$0.010 & ---          & 0.55$\pm$0.01 & ---           & \citet{Rozyczka09} \\
 8 & CoRoT 102932176         & 0.872241 &  7300 &  4547$\pm$80  & 0.405 & 0.3253$\pm$0.0001 & 0.1457$\pm$0.0001 & 0.027$\pm$0.012 & ---          & ---           & ---           & \citet{Lazaro15} \\
 9 & HI Mon                  & 1.57443  & 30000 & 29000$\pm$682 & 0.863 & 0.2815$\pm$0.0002 & 0.2739$\pm$0.0002 & 0.937$\pm$0.020 &7$\pm$2       & ---           & ---           & ASAS \\
10 & TYC 176-2950-1          & 11.55478 &  5735 &  5733$\pm$200 & 0.986 & 0.0601$\pm$0.0001 & 0.0430$\pm$0.0001 & 0.511$\pm$0.011 & ---          & ---           & ---           & ASAS \\
11 & CW CMa                  & 2.117977 &  9886 &  9639$\pm$200 & 1.058 & 0.1734$\pm$0.0002 & 0.1642$\pm$0.0002 & 0.994$\pm$0.013 & ---          & ---           & ---           & ASAS \\
12 & ASAS-08                 & 1.528489 &  4350 &  4310$\pm$205 & 0.978 & 0.1114$\pm$0.0001 & 0.1122$\pm$0.0001 & 0.944$\pm$0.020 & ---          & ---           & ---           & ASAS \\
13 & ASAS-09                 & 0.89742  &  4360 &  4360$\pm$150 & 1.004 & 0.1701$\pm$0.0002 & 0.1695$\pm$0.0002 & 0.993$\pm$0.012 & ---          & ---           & ---           & ASAS \\
14 & DU Leo                  & 1.374185 &  5940 &  5788$\pm$130 & 0.983 & 0.1916$\pm$0.0002 & 0.1969$\pm$0.0002 & 0.950$\pm$0.010 & ---          & ---           & ---           & ASAS \\
15 & QX Car                  & 4.47804  & 23800 & 22600$\pm$500 & 0.915 & 0.1442$\pm$0.0001 & 0.1361$\pm$0.0001 & 0.815$\pm$0.010 & ---          & 0.81$\pm$0.01 & ---           & \citet{Cousins81} \\
16 & HS Hya                  & 1.568041 &  6500 &  6329$\pm$50  & 0.971 & 0.1668$\pm$0.0001 & 0.1593$\pm$0.0001 & 0.818$\pm$0.010 & ---          & ---           & ---           & ASAS \\
17 & UW LMi                  & 3.874307 &  6500 &  6528$\pm$205 & 1.017 & 0.0956$\pm$0.0001 & 0.0940$\pm$0.0001 & 0.996$\pm$0.010 & ---          & ---           & ---           & {\it Hipparcos} \\
18 & eta Mus                 & 2.396316 & 12700 & 12550$\pm$300 & 0.999 & 0.1517$\pm$0.0001 & 0.1510$\pm$0.0002 & 0.960$\pm$0.008 & ---          & 0.93$\pm$0.01 & ---           & \citet{Bakis07} \\
19 & V1200 Cen               & 2.4828778&  6266 &  4706$\pm$205 & 0.621 & 0.1372$\pm$0.0002 & 0.1086$\pm$0.0001 & 0.168$\pm$0.003 & ---          & ---           & ---           & ASAS \\
20 & TV Nor                  & 8.524391 &  9120 &  7916$\pm$201 & 0.811 & 0.0685$\pm$0.0002 & 0.0578$\pm$0.0002 & 0.452$\pm$0.003 & ---          & 0.49$\pm$0.01 & ---           & \citet{North97}\\
21 & HD147827                & 8.87621  &  5957 &  5811$\pm$230 & 0.852 & 0.1100$\pm$0.0010 & 0.0830$\pm$0.0006 & 0.521$\pm$0.008 & ---          & ---           & ---           & ASAS \\
22 & FL Lyr                  & 2.178154 &  6150 &  5200$\pm$202 & 0.786 & 0.1398$\pm$0.0002 & 0.1049$\pm$0.0001 & 0.249$\pm$0.016 & ---          & ---           & ---           & \citet{Popper86} \\
23 & WOCS 24009              & 3.649303 &  5925 &  5860$\pm$201 & 0.987 & 0.0852$\pm$0.0001 & 0.0829$\pm$0.0001 & 0.975$\pm$0.011 &61.63$\pm$1.50& ---           & ---           & {\it Kepler} \\
24 & WOCS 40007              & 3.185096 &  6350 &  5930$\pm$150 & 0.875 & 0.1164$\pm$0.0002 & 0.0913$\pm$0.0002 & 0.505$\pm$0.013 & 0.43$\pm$0.03& 0.48$\pm$0.01 & 0.18$\pm$0.04 & \citet{Jeffries13} \\
25 & BD-20 5728              & 7.04335  &  5957 &  5943$\pm$205 & 0.721 & 0.1253$\pm$0.0002 & 0.0929$\pm$0.0002 & 0.542$\pm$0.017 & ---          & ---           & ---           & ASAS \\
26 & MT91 372                & 2.227582 & 24000 & 15616$\pm$200 & 0.440 & 0.2600$\pm$0.0003 & 0.1758$\pm$0.0003 & 0.210$\pm$0.022 & ---          & ---           & ---           & \citet{Kiminki15} \\
27 & MT91 696                & 1.469179 & 32000 & 30910$\pm$1050& 0.800 & 0.3902$\pm$0.0002 & 0.3509$\pm$0.0002 & 0.739$\pm$0.014 & ---          & ---           & ---           & \citet{Kiminki15} \\
28 & ASAS-21                 & 0.70243  &  4750 &  4220$\pm$180 & 0.851 & 0.2206$\pm$0.0001 & 0.1875$\pm$0.0001 & 0.418$\pm$0.012 & ---          & ---           & ---           & ASAS \\
29 & EK Cep                  & 4.427822 &  9000 &  5700$\pm$190 & 0.554 & 0.0949$\pm$0.0001 & 0.0791$\pm$0.0001 & 0.126$\pm$0.013 & ---          & 0.09$\pm$0.02 & ---           & \citet{Antonyuk09}\\
30 & OO Peg                  & 2.985    &  7850 &  7600$\pm$450 & 0.983 & 0.1648$\pm$0.0002 & 0.1462$\pm$0.0002 & 0.698$\pm$0.013 & ---          & ---           & ---           & ASAS \\
31 & NGC 7142 V2             & 15.6506  &  6238 &  6276$\pm$63  & 1.000 & 0.0438$\pm$0.0001 & 0.0437$\pm$0.0001 & 1.004$\pm$0.013 & ---          & 1.02$\pm$0.01 & ---           & \citet{Sandquist13} \\
32 & BG Ind                  & 1.464069 &  6353 &  6718$\pm$233 & 0.894 & 0.3073$\pm$0.0002 & 0.2254$\pm$0.0002 & 0.469$\pm$0.006 & ---          & ---           & ---           & ASAS \\
33 & V453 Cep                & 1.184725 & 10300 & 10410$\pm$500 & 0.962 & 0.2627$\pm$0.0002 & 0.2501$\pm$0.0002 & 1.063$\pm$0.005 & ---          & ---           & ---           & \citet{Griffin09} \\
\hline
 \end{tabular}%
\label{tab:addlabel}%
}
\end{table}%
\end{landscape}

\subsection{$BC$ coefficients}

The $BC$ coefficients computed from the parameters of 400 main-sequence stars (206 binaries) with metallicities $0.008\leq Z\leq 0.040$ in the Solar neighborhood are listed in Table 4. The columns of Table 4 are organized as: ID number, name of the system, $(B-V)_0$ for primary, $(B-V)_0$ for secondary, bolometric magnitude and its error for the primary, bolometric magnitude and its error for the secondary, $BC$ and its error for the primary, $BC$ and its error for the secondary. Notice that $BC$ values and associated uncertainties are not given for one of the components of the 12 systems, which are not on the main-sequence (Fig. 1b). Only bolometric absolute magnitudes are given for interested readers, who may want to calculate visual absolute magnitudes either using $BC$ values according to Eq. (1) or according to Eq. (5).   

\begin{table*}
{\scriptsize
  \centering
  \caption{De-reddened colours, bolometric magnitudes and bolometric correction coefficients of the present sample.}
    \begin{tabular}{clclcccccc}
\hline
 ID & Name & \multicolumn{1}{c}{Pri $(B-V)_0$} & \multicolumn{1}{c}{Sec $(B-V)_0$} & \multicolumn{1}{c}{Pri $M_{Bol}$} & \multicolumn{1}{c}{Sec $M_{Bol}$}& \multicolumn{1}{c}{Pri $BC$} & \multicolumn{1}{c}{Sec $BC$} \\
    &      & \multicolumn{1}{c}{(mag)}& \multicolumn{1}{c}{(mag)} & \multicolumn{1}{c}{(mag)} & \multicolumn{1}{c}{(mag)}& \multicolumn{1}{c}{(mag)} & \multicolumn{1}{c}{(mag)} \\
\hline
    1     & V421 Peg & ---   & ---   & 2.751$\pm$0.061 & 3.299$\pm$0.088 & 0.093$\pm$0.065 & 0.090$\pm$0.091 \\
    2     & DV Psc & 0.943 & 1.311 & 6.691$\pm$0.095 & 8.219$\pm$0.085 & -0.906$\pm$0.096 & -1.633$\pm$0.086 \\
    3     & MU Cas & -0.097 & -0.101 & -2.447$\pm$0.149 & -2.260$\pm$0.146 & -0.921$\pm$0.244 & -0.983$\pm$0.241 \\
    4     & GSC 4019-3345 & 0.114 & 0.118 & 1.781$\pm$0.168 & 1.781$\pm$0.294 & -0.142$\pm$0.209 & -0.125$\pm$0.319 \\
    5     & YZ Cas & -0.035 & 0.363 & 0.556$\pm$0.056 & 3.356$\pm$0.152 & -0.101$\pm$0.061 & 0.061$\pm$0.154 \\
    6     & NGC188 KR V12 & ---   & ---   & 3.876$\pm$0.079 & 3.973$\pm$0.080 & 0.178$\pm$0.117 & 0.177$\pm$0.118 \\
    7     & V364 Cas & 0.174 & 0.276 & 1.951$\pm$0.049 & 1.608$\pm$0.049 & -0.184$\pm$0.108 & -0.195$\pm$0.108 \\
    8     & zet Phe & -0.111 & -0.073 & -1.552$\pm$0.106 & 0.254$\pm$0.077 & -1.005$\pm$0.131 & -0.542$\pm$0.109 \\
    9     & CO And & ---   & ---   & 3.285$\pm$0.096 & 3.306$\pm$0.094 & -0.035$\pm$0.111 & -0.031$\pm$0.109 \\
    10    & V459 Cas & 0.070 & 0.076 & 1.228$\pm$0.143 & 1.295$\pm$0.144 & -0.094$\pm$0.165 & -0.093$\pm$0.165 \\
    11    & 2MASS J01132817-3821024 & ---   & ---   & 7.737$\pm$0.299 & 9.181$\pm$0.303 & -1.204$\pm$0.299 & -1.967$\pm$0.303 \\
    12    & UV Psc & 0.652 & 1.062 & 4.494$\pm$0.085 & 5.978$\pm$0.107 & -0.366$\pm$0.086 & -0.767$\pm$0.108 \\
    13    & NSVS 06507557 & ---   & ---   & 7.475$\pm$0.139 & ---   & -0.527$\pm$0.139 & --- \\
    14    & AL Ari & 0.428 & 0.690 & 3.672$\pm$0.056 & 5.246$\pm$0.065 & 0.167$\pm$0.068 & 0.048$\pm$0.076 \\
    15    & V615 Per & -0.163 & -0.075 & -1.209$\pm$0.197 & 0.541$\pm$0.225 & -1.304$\pm$0.298 & -0.193$\pm$0.317 \\
    16    & V505 Per & 0.396 & 0.425 & 3.667$\pm$0.136 & 3.736$\pm$0.137 & 0.075$\pm$0.137 & 0.090$\pm$0.138 \\
    17    & DN Cas & -0.208 & -0.203 & -7.004$\pm$0.136 & -5.935$\pm$0.172 & -3.535$\pm$0.201 & -3.171$\pm$0.226 \\
    18    & AG Ari & 0.006 & 0.037 & 0.481$\pm$0.108 & 0.857$\pm$0.104 & -0.150$\pm$0.126 & -0.045$\pm$0.123 \\
    19    & XY Cet & 0.179 & 0.231 & 2.027$\pm$0.075 & 2.286$\pm$0.080 & 0.229$\pm$0.089 & 0.195$\pm$0.093 \\
    20    & CW Eri & 0.314 & 0.379 & 2.399$\pm$0.075 & 3.205$\pm$0.117 & 0.006$\pm$0.079 & 0.019$\pm$0.120 \\
    21    & V799 Cas & -0.092 & -0.084 & -0.819$\pm$0.094 & -0.669$\pm$0.095 & -0.546$\pm$0.120 & -0.482$\pm$0.121 \\
    22    & AE For & 1.327 & 1.349 & 7.095$\pm$0.097 & 7.277$\pm$0.104 & -1.276$\pm$0.099 & -1.308$\pm$0.106 \\
    23    & V570 Per & 0.377 & 0.438 & 3.130$\pm$0.161 & 3.513$\pm$0.182 & 0.049$\pm$0.162 & 0.037$\pm$0.184 \\
    24    & IM Per & ---   & ---   & 1.647$\pm$0.087 & 1.692$\pm$0.093 & 0.266$\pm$0.117 & 0.263$\pm$0.121 \\
    25    & TV Cet & 0.346 & 0.414 & 3.106$\pm$0.099 & 3.741$\pm$0.101 & 0.049$\pm$0.101 & 0.025$\pm$0.103 \\
    26    & TZ For & 0.488 & ---   & 1.329$\pm$0.084 & ---   & -0.074$\pm$0.086 & --- \\
    27    & GJ 3236 & ---   & ---   & 9.239$\pm$0.148 & 9.988$\pm$0.148 & -2.537$\pm$0.157 & -2.511$\pm$0.156 \\
    28    & EY Cep & 0.315 & 0.336 & 3.021$\pm$0.093 & 3.087$\pm$0.095 & 0.009$\pm$0.097 & 0.013$\pm$0.099 \\
    29    & V1229 Tau & -0.064 & 0.212 & 1.189$\pm$0.135 & 2.638$\pm$0.183 & -0.032$\pm$0.140 & 0.243$\pm$0.186 \\
    30    & IQ Per & -0.105 & 0.178 & -0.505$\pm$0.075 & 2.612$\pm$0.064 & -0.491$\pm$0.089 & 0.295$\pm$0.079 \\
    31    & SZ Cam & -0.226 & -0.195 & -7.333$\pm$0.038 & -5.902$\pm$0.058 & -3.052$\pm$0.219 & -2.557$\pm$0.223 \\
    32    & V1094 Tau & 0.684 & 0.724 & 3.934$\pm$0.075 & 4.575$\pm$0.077 & -0.056$\pm$0.091 & -0.060$\pm$0.094 \\
    33    & NP Per & ---   & ---   & 3.591$\pm$0.064 & 5.335$\pm$0.155 & -0.033$\pm$0.087 & -0.457$\pm$0.166 \\
    34    & V818 Tau & 0.710 & 1.193 & 5.207$\pm$0.457 & 6.934$\pm$0.281 & 0.122$\pm$0.458 & -0.444$\pm$0.282 \\
    35    & AL Dor & 0.568 & 0.556 & 4.374$\pm$0.082 & 4.339$\pm$0.082 & -0.016$\pm$0.084 & -0.011$\pm$0.084 \\
    36    & TYC 4749-560-1 & ---   & ---   & 5.436$\pm$0.163 & 5.653$\pm$0.170 & -0.289$\pm$0.207 & -0.149$\pm$0.213 \\
    37    & HP Aur & ---   & ---   & 4.652$\pm$0.090 & 5.778$\pm$0.101 & -0.210$\pm$0.097 & -0.327$\pm$0.108 \\
    38    & V1236 Tau  & ---   & ---   & 6.700$\pm$0.211 & 6.667$\pm$0.264 & -0.461$\pm$0.441 & -0.453$\pm$0.469 \\
    39    & CD Tau & 0.433 & 0.436 & 3.155$\pm$0.041 & 3.431$\pm$0.044 & 0.174$\pm$0.044 & 0.169$\pm$0.047 \\
    40    & AR Aur & -0.073 & -0.040 & 0.654$\pm$0.120 & 0.915$\pm$0.126 & -0.228$\pm$0.124 & -0.138$\pm$0.130 \\
    41    & EW Ori & ---   & ---   & 4.182$\pm$0.069 & 4.442$\pm$0.071 & -0.010$\pm$0.074 & -0.070$\pm$0.076 \\
    42    & AS Cam & ---   & ---   & -0.528$\pm$0.220 & 0.562$\pm$0.216 & -0.608$\pm$0.232 & -0.515$\pm$0.228 \\
    43    & UX Men & 0.500 & 0.551 & 3.779$\pm$0.073 & 3.937$\pm$0.074 & 0.013$\pm$0.075 & 0.022$\pm$0.076 \\
    44    & TZ Men & -0.114 & 0.258 & 0.659$\pm$0.210 & 2.999$\pm$0.182 & -0.287$\pm$0.219 & 0.076$\pm$0.192 \\
    45    & V432 Aur & ---   & 0.389 & ---   & 3.649$\pm$0.012 & ---   & -0.164$\pm$0.026 \\
    46    & GG Ori & -0.048 & -0.053 & 1.037$\pm$0.092 & 1.063$\pm$0.092 & 0.025$\pm$0.159 & 0.029$\pm$0.159 \\
    47    & beta Aur & 0.044 & 0.012 & 0.439$\pm$0.094 & 0.667$\pm$0.095 & -0.173$\pm$0.095 & -0.009$\pm$0.096 \\
    48    & V530 Ori & 0.583 & 1.337 & 4.696$\pm$0.079 & 7.621$\pm$0.137 & -0.021$\pm$0.083 & -1.061$\pm$0.139 \\
    49    & V1388 Ori & ---   & ---   & -4.507$\pm$0.107 & -3.289$\pm$0.116 & -2.214$\pm$0.366 & -2.154$\pm$0.369 \\
    50    & FT Ori & 0.006 & 0.117 & 1.170$\pm$0.182 & 1.953$\pm$0.152 & -0.074$\pm$0.193 & 0.052$\pm$0.166 \\
    51    & V404 CMa & 0.877 & 1.005 & 6.831$\pm$0.112 & 7.229$\pm$0.058 & -0.467$\pm$0.232 & -0.587$\pm$0.212 \\
    52    & IM Mon & -0.212 & -0.176 & -2.569$\pm$0.091 & -1.126$\pm$0.167 & -1.373$\pm$0.116 & -0.814$\pm$0.182 \\
    53    & RR Lyn & 0.193 & 0.260 & 1.524$\pm$0.060 & 2.919$\pm$0.075 & 0.136$\pm$0.115 & 0.135$\pm$0.123 \\
    54    & KL CMa & ---   & ---   & -0.726$\pm$0.089 & 0.320$\pm$0.248 & -0.313$\pm$0.110 & -0.403$\pm$0.257 \\
    55    & V578 Mon & -0.240 & -0.210 & -6.084$\pm$0.074 & -4.917$\pm$0.078 & -2.440$\pm$0.715 & -2.107$\pm$0.715 \\
    56    & WW Aur & 0.143 & 0.163 & 1.919$\pm$0.229 & 2.179$\pm$0.233 & 0.064$\pm$0.238 & 0.066$\pm$0.241 \\
    57    & TYC 7091-888-1 & ---   & 0.734 & ---   & 5.716$\pm$0.069 & ---   & -0.133$\pm$0.082 \\
    58    & V501 Mon & 0.183 & 0.276 & 2.217$\pm$0.067 & 2.893$\pm$0.068 & 0.099$\pm$0.125 & 0.109$\pm$0.125 \\
    59    & GX Gem & 0.509 & 0.521 & 2.593$\pm$0.071 & 2.698$\pm$0.071 & 0.250$\pm$0.137 & 0.234$\pm$0.138 \\
    60    & CoRoT 102932176 & ---   & ---   & 2.659$\pm$0.073 & 6.459$\pm$0.097 & 0.009$\pm$0.154 & -0.099$\pm$0.166 \\
    61    & HS Aur & ---   & ---   & 5.059$\pm$0.075 & 5.486$\pm$0.096 & 0.084$\pm$0.077 & 0.052$\pm$0.097 \\
    62    & HI Mon & ---   & ---   & -5.813$\pm$0.086 & -5.606$\pm$0.107 & -2.738$\pm$0.525 & -2.601$\pm$0.529 \\
    63    & LT CMa & ---   & ---   & -2.733$\pm$0.135 & -0.387$\pm$0.270 & -1.558$\pm$0.161 & -0.857$\pm$0.284 \\
    64    & SW CMa & 0.139 & 0.148 & 0.819$\pm$0.081 & 1.282$\pm$0.088 & 0.270$\pm$0.107 & 0.306$\pm$0.113 \\
    65    & GZ CMa & 0.073 & 0.111 & 0.923$\pm$0.175 & 1.401$\pm$0.178 & -0.066$\pm$0.178 & -0.019$\pm$0.182 \\
    66    & TYC 176-2950-1 & ---   & ---   & ---   & 4.406$\pm$0.156 & ---   & 0.093$\pm$0.163 \\
    67    & CW CMa & ---   & ---   & 1.002$\pm$0.099 & 1.231$\pm$0.124 & -0.451$\pm$0.105 & -0.229$\pm$0.129 \\
    68    & FS Mon & 0.362 & 0.395 & 2.512$\pm$0.066 & 3.121$\pm$0.068 & 0.049$\pm$0.073 & 0.032$\pm$0.075 \\
    69    & YY Gem & ---   & ---   & 7.567$\pm$0.116 & 7.624$\pm$0.116 & -1.547$\pm$0.116 & -1.508$\pm$0.116 \\
    70    & V392 Car & ---   & ---   & 1.831$\pm$0.106 & 1.963$\pm$0.109 & 0.117$\pm$0.113 & 0.131$\pm$0.115 \\
    71    & AI Hya & ---   & ---   & 1.633$\pm$0.040 & 1.131$\pm$0.042 & 0.445$\pm$0.098 & 0.321$\pm$0.099 \\
    72    & ASAS-08 & ---   & ---   & 6.762$\pm$0.207 & 6.786$\pm$0.210 & -1.349$\pm$0.218 & -1.387$\pm$0.222 \\
    73    & AY Cam & 0.398 & 0.381 & 1.527$\pm$0.062 & 2.122$\pm$0.061 & 0.051$\pm$0.068 & 0.040$\pm$0.067 \\
    74    & HD 71636 & 0.389 & 0.501 & 2.946$\pm$0.088 & 3.589$\pm$0.095 & 0.054$\pm$0.090 & 0.036$\pm$0.097 \\
    75    & CU Cnc & ---   & ---   & 9.174$\pm$0.208 & 9.438$\pm$0.214 & -2.201$\pm$0.208 & -2.293$\pm$0.215 \\
    \end{tabular}%
  \label{tab:addlabel}%
}
\end{table*}%
 
\begin{table*}
\setcounter{table}{4} 
\contcaption{}
{\scriptsize
  \centering
    \begin{tabular}{clcccccc}
\hline
 ID & Name & \multicolumn{1}{c}{Pri $(B-V)_0$} & \multicolumn{1}{c}{Sec $(B-V)_0$} & \multicolumn{1}{c}{Pri $M_{Bol}$} & \multicolumn{1}{c}{Sec $M_{Bol}$}& \multicolumn{1}{c}{Pri $BC$} & \multicolumn{1}{c}{Sec $BC$} \\
    &      & \multicolumn{1}{c}{(mag)}& \multicolumn{1}{c}{(mag)} & \multicolumn{1}{c}{(mag)} & \multicolumn{1}{c}{(mag)}& \multicolumn{1}{c}{(mag)} & \multicolumn{1}{c}{(mag)} \\
\hline
    76    & VZ Hya & 0.423 & 0.508 & 3.534$\pm$0.098 & 4.134$\pm$0.104 & -0.025$\pm$0.103 & -0.038$\pm$0.109 \\
    77    & KX Cnc & 0.563 & 0.578 & 4.401$\pm$0.079 & 4.471$\pm$0.080 & -0.017$\pm$0.081 & -0.022$\pm$0.082 \\
    78    & RS Cha & 0.182 & 0.186 & 1.861$\pm$0.074 & 1.899$\pm$0.070 & 0.120$\pm$0.076 & 0.174$\pm$0.071 \\
    79    & V467 Vel & -0.344 & -0.299 & -8.232$\pm$0.022 & -4.385$\pm$0.023 & -2.946$\pm$0.510 & -2.126$\pm$0.511 \\
    80    & NSVS 02502726 & 0.998 & ---   & 6.777$\pm$0.203 & ---   & -0.351$\pm$0.204 & --- \\
    81    & EPIC 211409263 & 0.565 & ---   & 4.351$\pm$0.073 & ---   & 0.134$\pm$0.091 & --- \\
    82    & CV Vel & -0.198 & -0.196 & -3.253$\pm$0.122 & -3.124$\pm$0.123 & -1.466$\pm$0.133 & -1.389$\pm$0.134 \\
    83    & XY UMa & 0.753 & 0.957 & 4.869$\pm$0.038 & 7.202$\pm$0.035 & -0.517$\pm$0.039 & -1.050$\pm$0.037 \\
    84    & PT Vel & -0.084 & 0.145 & 1.087$\pm$0.073 & 2.558$\pm$0.106 & 0.044$\pm$0.166 & 0.190$\pm$0.182 \\
    85    & KW Hya & 0.234 & 0.432 & 1.681$\pm$0.110 & 3.105$\pm$0.130 & 0.014$\pm$0.112 & 0.015$\pm$0.132 \\
    86    & ASAS-09 & ---   & ---   & 6.529$\pm$0.153 & 6.537$\pm$0.154 & -0.448$\pm$0.382 & -0.447$\pm$0.383 \\
    87    & DU Leo & ---   & ---   & 4.252$\pm$0.092 & 4.365$\pm$0.112 & 0.058$\pm$0.093 & 0.114$\pm$0.114 \\
    88    & QX Car & -0.224 & -0.218 & -4.576$\pm$0.096 & -4.227$\pm$0.101 & -2.340$\pm$0.138 & -2.213$\pm$0.142 \\
    89    & HS Hya & ---   & ---   & 3.691$\pm$0.035 & 3.908$\pm$0.037 & 0.126$\pm$0.041 & 0.125$\pm$0.042 \\
    90    & UV Leo & 0.597 & 0.623 & 4.573$\pm$0.072 & 4.373$\pm$0.088 & -0.220$\pm$0.073 & -0.276$\pm$0.089 \\
    91    & RZ Cha & 0.395 & 0.406 & 2.461$\pm$0.109 & 2.461$\pm$0.109 & -0.078$\pm$0.111 & 0.025$\pm$0.111 \\
    92    & DW Car & -0.213 & -0.208 & -5.398$\pm$0.157 & -5.046$\pm$0.166 & -2.148$\pm$0.281 & -1.987$\pm$0.286 \\
    93    & UW LMi & ---   & ---   & 3.707$\pm$0.163 & 3.724$\pm$0.174 & -0.223$\pm$0.164 & -0.210$\pm$0.175 \\
    94    & EM Car & -0.255 & -0.256 & -7.817$\pm$0.258 & -7.569$\pm$0.259 & -2.709$\pm$0.416 & -2.709$\pm$0.416 \\
    95    & FM Leo & 0.463 & 0.498 & 3.263$\pm$0.174 & 3.539$\pm$0.163 & 0.106$\pm$0.177 & 0.095$\pm$0.166 \\
    96    & MW UMa & ---   & ---   & 3.716$\pm$0.088 & 4.211$\pm$0.025 & -0.142$\pm$0.092 & -0.051$\pm$0.036 \\
    97    & EP Cru & -0.158 & -0.152 & -2.381$\pm$0.140 & -2.239$\pm$0.143 & -1.110$\pm$0.178 & -1.083$\pm$0.181 \\
    98    & VV Crv & 0.423 & 0.413 & 1.583$\pm$0.134 & 3.045$\pm$0.131 & 0.100$\pm$0.135 & 0.093$\pm$0.132 \\
    99    & IM Vir & 0.621 & 1.106 & 4.764$\pm$0.085 & 6.904$\pm$0.139 & -0.249$\pm$0.086 & -0.921$\pm$0.140 \\
    100   & HY Vir & 0.384 & 0.459 & 1.734$\pm$0.079 & 3.265$\pm$0.080 & -0.010$\pm$0.086 & -0.034$\pm$0.087 \\
    101   & eta Mus & -0.132 & -0.089 & -0.337$\pm$0.040 & -0.275$\pm$0.112 & -0.388$\pm$0.063 & -0.371$\pm$0.122 \\
    102   & SZ Cen & 0.120 & 0.147 & 0.666$\pm$0.113 & 0.297$\pm$0.116 & 0.151$\pm$0.126 & 0.159$\pm$0.128 \\
    103   & V1200 Cen & ---   & ---   & 3.668$\pm$0.274 & 5.420$\pm$0.529 & 0.125$\pm$0.276 & -0.062$\pm$0.530 \\
    104   & ZZ Boo & 0.373 & 0.329 & 2.314$\pm$0.071 & 2.270$\pm$0.071 & -0.046$\pm$0.073 & -0.090$\pm$0.073 \\
    105   & BH Vir & 0.467 & 0.555 & 4.054$\pm$0.114 & 4.707$\pm$0.176 & -0.139$\pm$0.118 & -0.169$\pm$0.179 \\
    106   & DM Vir & 0.469 & 0.469 & 2.992$\pm$0.070 & 2.992$\pm$0.202 & -0.002$\pm$0.076 & -0.002$\pm$0.204 \\
    107   & AD Boo & 0.377 & 0.486 & 3.135$\pm$0.081 & 4.040$\pm$0.087 & -0.014$\pm$0.093 & -0.097$\pm$0.097 \\
    108   & GG Lup & -0.149 & -0.076 & -0.705$\pm$0.110 & 0.836$\pm$0.132 & -0.726$\pm$0.123 & -0.408$\pm$0.143 \\
    109   & AQ Ser & ---   & ---   & 2.481$\pm$0.069 & 2.386$\pm$0.073 & 0.152$\pm$0.087 & 0.149$\pm$0.090 \\
    110   & CV Boo & ---   & ---   & 4.232$\pm$0.120 & 4.458$\pm$0.122 & -0.173$\pm$0.137 & -0.055$\pm$0.140 \\
    111   & V335 Ser & 0.049 & 0.088 & 1.026$\pm$0.134 & 1.843$\pm$0.122 & -0.167$\pm$0.136 & -0.095$\pm$0.125 \\
    112   & TV Nor & 0.062 & -0.029 & 1.416$\pm$0.072 & 2.403$\pm$0.112 & -0.049$\pm$0.084 & 0.075$\pm$0.120 \\
    113   & V760 Sco & -0.164 & -0.155 & -2.331$\pm$0.136 & -1.890$\pm$0.139 & -1.467$\pm$0.151 & -1.378$\pm$0.154 \\
    114   & HD147827 & ---   & ---   & 2.569$\pm$0.318 & ---   & 0.151$\pm$0.321 & --- \\
    115   & V349 Ara & ---   & ---   & -0.006$\pm$0.105 & 0.135$\pm$0.113 & -0.177$\pm$0.124 & -0.005$\pm$0.131 \\
    116   & V923 Sco & 0.334 & 0.367 & 2.535$\pm$0.064 & 2.826$\pm$0.066 & 0.099$\pm$0.066 & 0.083$\pm$0.067 \\
    117   & V2626 Oph & 0.329 & 0.482 & 0.903$\pm$0.065 & 3.226$\pm$0.068 & -0.033$\pm$0.073 & -0.096$\pm$0.075 \\
    118   & WZ Oph & 0.503 & 0.509 & 3.722$\pm$0.073 & 3.729$\pm$0.073 & -0.082$\pm$0.078 & -0.062$\pm$0.078 \\
    119   & V2365 Oph & 0.021 & 0.532 & 0.873$\pm$0.092 & 4.440$\pm$0.143 & -0.324$\pm$0.113 & -0.179$\pm$0.157 \\
    120   & V2368 Oph & 0.114 & 0.103 & -0.263$\pm$0.094 & -0.288$\pm$0.092 & -0.231$\pm$0.099 & -0.269$\pm$0.097 \\
    121   & U Oph & -0.248 & -0.164 & -2.481$\pm$0.387 & -2.050$\pm$0.409 & -1.800$\pm$0.471 & -1.595$\pm$0.489 \\
    122   & TX Her & 0.222 & 0.317 & 2.446$\pm$0.122 & 3.333$\pm$0.145 & 0.098$\pm$0.123 & 0.180$\pm$0.146 \\
    123   & LV Her & 0.573 & 0.581 & 3.864$\pm$0.109 & 3.959$\pm$0.110 & 0.029$\pm$0.128 & 0.028$\pm$0.128 \\
    124   & V501 Her & 0.719 & 0.703 & 3.301$\pm$0.076 & 3.883$\pm$0.076 & -0.032$\pm$0.093 & -0.015$\pm$0.092 \\
    125   & V624 Her & ---   & ---   & 0.836$\pm$0.083 & 1.632$\pm$0.087 & 0.103$\pm$0.085 & 0.107$\pm$0.089 \\
    126   & BD-00 3357 & ---   & ---   & 2.498$\pm$0.245 & 3.672$\pm$0.117 & 0.160$\pm$0.252 & -0.124$\pm$0.131 \\
    127   & V539 Ara & -0.133 & -0.264 & -3.482$\pm$0.316 & -2.808$\pm$0.343 & -1.994$\pm$0.443 & -1.963$\pm$0.463 \\
    128   & V2653 Oph & ---   & ---   & 2.207$\pm$0.305 & 2.820$\pm$0.449 & -0.318$\pm$0.394 & -0.282$\pm$0.514 \\
    129   & Z Her & 0.338 & ---   & 2.947$\pm$0.087 & ---   & 0.020$\pm$0.087 & --- \\
    130   & V994 Her-B & 0.008 & 0.095 & 1.813$\pm$0.163 & 2.214$\pm$0.122 & -0.309$\pm$0.443 & -0.106$\pm$0.430 \\
    131   & V994 Her-A & -0.080 & 0.000 & -0.073$\pm$0.103 & 1.462$\pm$0.065 & -0.698$\pm$0.425 & -0.200$\pm$0.417 \\
    132   & V451 Oph & -0.122 & -0.070 & -0.094$\pm$0.323 & 0.901$\pm$0.224 & -0.561$\pm$0.329 & -0.336$\pm$0.233 \\
    133   & RX Her & ---   & ---   & -0.053$\pm$0.102 & 0.843$\pm$0.125 & -0.884$\pm$0.106 & -0.538$\pm$0.128 \\
    134   & V413 Ser & ---   & ---   & -0.626$\pm$0.122 & -0.124$\pm$0.123 & -0.346$\pm$0.588 & -0.075$\pm$0.588 \\
    135   & QY Tel & ---   & ---   & 2.988$\pm$0.229 & 2.207$\pm$0.189 & -0.235$\pm$0.237 & -0.220$\pm$0.199 \\
    136   & V4403 Sgr & ---   & ---   & 3.010$\pm$0.103 & 2.359$\pm$0.107 & 0.008$\pm$0.107 & 0.034$\pm$0.110 \\
    137   & V1331 Aql & -0.255 & -0.184 & -4.839$\pm$0.023 & -3.712$\pm$0.034 & -2.568$\pm$0.086 & -2.048$\pm$0.090 \\
    138   & YY Sgr & -0.154 & -0.141 & -1.404$\pm$0.207 & -0.996$\pm$0.211 & -0.973$\pm$0.254 & -0.880$\pm$0.257 \\
    139   & DI Her & -0.180 & -0.153 & -2.086$\pm$0.209 & -1.418$\pm$0.210 & -1.270$\pm$0.216 & -0.979$\pm$0.217 \\
    140   & HP Dra & 0.462 & 0.719 & 3.936$\pm$0.110 & 4.588$\pm$0.113 & -0.083$\pm$0.111 & -0.105$\pm$0.113 \\
    141   & V1182 Aql & -0.292 & -0.243 & -8.753$\pm$0.067 & -5.923$\pm$0.107 & -4.378$\pm$0.162 & -3.340$\pm$0.183 \\
    142   & V1665 Aql & -0.097 & -0.074 & -1.626$\pm$0.134 & -0.386$\pm$0.126 & -0.860$\pm$0.149 & -0.728$\pm$0.141 \\
    143   & V805 Aql & 0.152 & 0.326 & 1.589$\pm$0.214 & 2.566$\pm$0.235 & -0.210$\pm$0.216 & -0.138$\pm$0.236 \\
    144   & FL Lyr & ---   & ---   & 3.925$\pm$0.151 & 5.277$\pm$0.182 & 0.019$\pm$0.153 & -0.139$\pm$0.183 \\
    145   & V565 Lyr & ---   & ---   & 4.662$\pm$0.075 & 5.069$\pm$0.102 & 0.159$\pm$0.993 & 0.072$\pm$0.996 \\
    146   & V1430 Aql & 0.790 & 0.695 & 4.933$\pm$0.125 & 5.775$\pm$0.092 & -0.128$\pm$0.128 & -0.154$\pm$0.095 \\
    147   & UZ Dra & 0.444 & 0.499 & 3.834$\pm$0.092 & 4.279$\pm$0.088 & 0.114$\pm$0.094 & 0.068$\pm$0.091 \\
    148   & V2080 Cyg & 0.500 & 0.505 & 3.557$\pm$0.055 & 3.562$\pm$0.055 & 0.067$\pm$0.057 & 0.068$\pm$0.057 \\
    149   & V2083 Cyg & 0.223 & 0.216 & 1.726$\pm$0.176 & 1.416$\pm$0.179 & 0.346$\pm$0.336 & 0.320$\pm$0.338 \\
    150   & V885 Cyg & ---   & ---   & 1.145$\pm$0.079 & 0.466$\pm$0.082 & 0.163$\pm$0.109 & 0.175$\pm$0.112 \\
    \end{tabular}%
  \label{tab:addlabel}%
}
\end{table*}%

\begin{table*}
\setcounter{table}{4} 
\contcaption{}
{\scriptsize
  \centering
    \begin{tabular}{clclcccccc}
\hline
 ID & Name & \multicolumn{1}{c}{Pri $(B-V)_0$} & \multicolumn{1}{c}{Sec $(B-V)_0$} & \multicolumn{1}{c}{Pri $M_{Bol}$} & \multicolumn{1}{c}{Sec $M_{Bol}$}& \multicolumn{1}{c}{Pri $BC$} & \multicolumn{1}{c}{Sec $BC$} \\
    &      & \multicolumn{1}{c}{(mag)}& \multicolumn{1}{c}{(mag)} & \multicolumn{1}{c}{(mag)} & \multicolumn{1}{c}{(mag)}& \multicolumn{1}{c}{(mag)} & \multicolumn{1}{c}{(mag)} \\
\hline
    151   & V4089 Sgr & 0.029 & 0.243 & 0.106$\pm$0.052 & 2.657$\pm$0.066 & 0.114$\pm$0.063 & 0.109$\pm$0.074 \\
    152   & KIC 9777062 & 0.275 & 0.355 & 2.281$\pm$0.141 & 2.894$\pm$0.156 & 0.064$\pm$0.158 & 0.087$\pm$0.171 \\
    153   & V1143 Cyg & 0.443 & 0.456 & 3.604$\pm$0.077 & 3.682$\pm$0.078 & 0.063$\pm$0.078 & 0.066$\pm$0.079 \\
    154   & WOCS 24009 & ---   & ---   & 4.421$\pm$0.148 & 4.529$\pm$0.151 & 0.089$\pm$0.248 & 0.169$\pm$0.250 \\
    155   & WOCS 40007 & 0.543 & 0.597 & 3.596$\pm$0.103 & 4.420$\pm$0.110 & 0.260$\pm$0.243 & 0.341$\pm$0.246 \\
    156   & V541 Cyg & -0.145 & -0.143 & 0.733$\pm$0.083 & 0.918$\pm$0.086 & -0.453$\pm$0.122 & -0.333$\pm$0.124 \\
    157   & V1765 Cyg & -0.236 & -0.276 & -8.810$\pm$0.436 & -5.514$\pm$0.438 & -3.162$\pm$0.458 & -2.642$\pm$0.460 \\
    158   & V380 Cyg & ---   & -0.227 & ---   & -4.117$\pm$0.231 & ---   & -2.273$\pm$0.315 \\
    159   & BD-20 5728 & ---   & ---   & 2.413$\pm$0.182 & ---   & -0.001$\pm$0.187 & --- \\
    160   & BS Dra & 0.284 & 0.349 & 3.351$\pm$0.106 & 3.375$\pm$0.149 & 0.225$\pm$0.108 & 0.092$\pm$0.151 \\
    161   & V477 Cyg & 0.096 & 0.375 & 1.965$\pm$0.165 & 3.560$\pm$0.167 & -0.024$\pm$0.166 & 0.186$\pm$0.168 \\
    162   & V453 Cyg & -0.270 & -0.271 & -6.559$\pm$0.083 & -5.413$\pm$0.139 & -3.069$\pm$0.152 & -2.989$\pm$0.188 \\
    163   & V478 Cyg & -0.329 & -0.359 & -6.866$\pm$0.156 & -6.866$\pm$0.156 & -2.332$\pm$0.241 & -2.524$\pm$0.241 \\
    164   & MY Cyg & 0.255 & 0.266 & 2.145$\pm$0.125 & 2.117$\pm$0.126 & 0.094$\pm$0.127 & 0.090$\pm$0.128 \\
    165   & V399 Vul & ---   & ---   & -4.498$\pm$0.074 & -2.983$\pm$0.135 & -2.475$\pm$0.129 & -1.538$\pm$0.172 \\
    166   & BP Vul & 0.209 & 0.329 & 2.144$\pm$0.086 & 3.149$\pm$0.098 & 0.085$\pm$0.118 & 0.191$\pm$0.127 \\
    167   & V442 Cyg & 0.401 & 0.434 & 2.370$\pm$0.070 & 2.941$\pm$0.075 & 0.029$\pm$0.078 & 0.021$\pm$0.083 \\
    168   & MP Del & 0.268 & 0.398 & 1.734$\pm$0.082 & 2.960$\pm$0.086 & 0.309$\pm$0.084 & 0.278$\pm$0.089 \\
    169   & V456 Cyg & 0.238 & 0.343 & 2.334$\pm$0.062 & 3.220$\pm$0.259 & 0.080$\pm$0.087 & 0.188$\pm$0.266 \\
    170   & MT91 372 & ---   & ---   & -4.917$\pm$0.040 & -2.201$\pm$0.068 & -1.751$\pm$0.186 & -0.731$\pm$0.194 \\
    171   & MT91 696 & ---   & ---   & -6.716$\pm$0.082 & -6.335$\pm$0.148 & -2.872$\pm$0.237 & -2.820$\pm$0.267 \\
    172   & IO Aqr & ---   & ---   & 2.538$\pm$0.094 & 2.352$\pm$0.087 & -0.021$\pm$0.097 & -0.029$\pm$0.091 \\
    173   & V379 Cep & -0.243 & -0.207 & -5.588$\pm$0.091 & -3.137$\pm$0.085 & -1.793$\pm$0.105 & -1.577$\pm$0.101 \\
    174   & Y Cyg & -0.288 & -0.288 & -6.670$\pm$0.043 & -6.723$\pm$0.024 & -2.971$\pm$0.170 & -3.007$\pm$0.166 \\
    175   & CG Cyg & 0.738 & 0.914 & 5.382$\pm$0.156 & 5.977$\pm$0.071 & -0.031$\pm$0.159 & -0.511$\pm$0.077 \\
    176   & V1061 Cyg & 0.521 & 0.747 & 3.401$\pm$0.074 & 5.168$\pm$0.131 & 0.047$\pm$0.076 & -0.167$\pm$0.132 \\
    177   & EI Cep & 0.345 & 0.304 & 1.750$\pm$0.074 & 2.097$\pm$0.075 & 0.149$\pm$0.077 & 0.181$\pm$0.078 \\
    178   & ASAS-21 & ---   & ---   & 5.947$\pm$0.141 & 6.813$\pm$0.192 & -0.655$\pm$0.149 & -0.735$\pm$0.199 \\
    179   & EK Cep & -0.044 & 0.374 & 1.819$\pm$0.099 & 4.200$\pm$0.147 & 0.187$\pm$0.224 & 0.322$\pm$0.249 \\
    180   & OO Peg & ---   & ---   & 1.683$\pm$0.194 & 2.084$\pm$0.257 & -0.106$\pm$0.196 & -0.095$\pm$0.259 \\
    181   & NGC 7142 V2 & 0.468 & 0.456 & 3.361$\pm$0.037 & 3.338$\pm$0.044 & 0.329$\pm$0.165 & 0.311$\pm$0.167 \\
    182   & VZ Cep & 0.456 & 0.596 & 3.042$\pm$0.032 & 4.996$\pm$0.063 & -0.007$\pm$0.054 & -0.033$\pm$0.076 \\
    183   & V497 Cep & -0.203 & -0.189 & -3.396$\pm$0.091 & -2.480$\pm$0.102 & -1.717$\pm$0.121 & -1.479$\pm$0.129 \\
    184   & BG Ind & ---   & ---   & 2.490$\pm$0.185 & 2.921$\pm$0.159 & 0.091$\pm$0.204 & -0.301$\pm$0.181 \\
    185   & CM Lac & 0.051 & 0.232 & 2.065$\pm$0.156 & 2.931$\pm$0.172 & 0.256$\pm$0.157 & 0.369$\pm$0.173 \\
    186   & V398 Lac & ---   & ---   & -1.509$\pm$0.213 & 0.031$\pm$0.204 & -0.401$\pm$0.228 & -0.365$\pm$0.220 \\
    187   & BW Aqr & 0.488 & 0.479 & 2.759$\pm$0.080 & 2.996$\pm$0.083 & -0.001$\pm$0.143 & -0.062$\pm$0.144 \\
    188   & WX Cep & 0.107 & 0.085 & 0.233$\pm$0.134 & 0.710$\pm$0.112 & -0.164$\pm$0.140 & -0.105$\pm$0.119 \\
    189   & LL Aqr & 0.530 & 0.624 & 3.910$\pm$0.034 & 4.788$\pm$0.040 & 0.046$\pm$0.039 & -0.025$\pm$0.044 \\
    190   & RW Lac & ---   & ---   & 4.366$\pm$0.076 & 4.969$\pm$0.118 & -0.106$\pm$0.077 & -0.106$\pm$0.119 \\
    191   & AH Cep & ---   & ---   & -6.429$\pm$0.150 & -6.051$\pm$0.159 & -2.788$\pm$0.165 & -2.725$\pm$0.173 \\
    192   & V364 Lac & 0.156 & 0.160 & 0.591$\pm$0.083 & 0.684$\pm$0.081 & -0.116$\pm$0.095 & -0.136$\pm$0.093 \\
    193   & V453 Cep & ---   & ---   & 0.601$\pm$0.295 & 0.662$\pm$0.301 & -0.391$\pm$0.492 & -0.265$\pm$0.496 \\
    194   & EF Aqr & 0.525 & 0.823 & 3.832$\pm$0.050 & 5.303$\pm$0.096 & 0.017$\pm$0.059 & -0.151$\pm$0.101 \\
    195   & CW Cep & -0.247 & -0.239 & -5.876$\pm$0.161 & -5.579$\pm$0.165 & -2.344$\pm$1.026 & -2.212$\pm$1.027 \\
    196   & PV Cas & 0.035 & 0.035 & 0.505$\pm$0.108 & 0.458$\pm$0.108 & -0.270$\pm$0.128 & -0.265$\pm$0.129 \\
    197   & RT And & 0.458 & 0.633 & 3.988$\pm$0.155 & 5.817$\pm$0.106 & -0.190$\pm$0.156 & -0.325$\pm$0.107 \\
    198   & V396 Cas & 0.037 & 0.135 & 0.635$\pm$0.071 & 1.783$\pm$0.062 & -0.140$\pm$0.085 & -0.069$\pm$0.078 \\
    199   & NSVS 11868841 & ---   & ---   & 5.130$\pm$0.105 & ---   & 0.044$\pm$0.111 & --- \\
    200   & AR Cas & -0.246 & 0.181 & -3.333$\pm$0.068 & 2.182$\pm$0.067 & -1.701$\pm$0.108 & 0.021$\pm$0.107 \\
    201   & V731 Cep & -0.073 & 0.016 & 0.755$\pm$0.089 & 1.511$\pm$0.108 & -0.422$\pm$0.139 & -0.198$\pm$0.152 \\
    202   & IT Cas & 0.434 & 0.430 & 3.220$\pm$0.077 & 3.266$\pm$0.092 & -0.005$\pm$0.099 & 0.003$\pm$0.111 \\
    203   & BK Peg & 0.513 & 0.502 & 2.895$\pm$0.060 & 3.507$\pm$0.032 & 0.090$\pm$0.071 & 0.095$\pm$0.050 \\
    204   & AP And & 0.442 & 0.448 & 3.725$\pm$0.100 & 3.840$\pm$0.101 & -0.057$\pm$0.136 & -0.068$\pm$0.137 \\
    205   & AL Scl & -0.129 & 0.125 & -1.520$\pm$0.117 & 1.493$\pm$0.155 & -0.696$\pm$0.134 & -0.881$\pm$0.168 \\
    206   & V821 Cas & 0.022 & 0.135 & 0.806$\pm$0.187 & 2.293$\pm$0.205 & -0.146$\pm$0.189 & -0.029$\pm$0.207 \\
\hline
    \end{tabular}%
  \label{tab:addlabel}%
}
\end{table*}%

\subsection{$(B-V)_0$ colours}

$(B-V)_0$ colours in Table 4 are just by-products only for the systems if their light ratio ($L_2/L_1$) in the $B$ band is available together with the light ratio in the $V$ band. According to the standard extinction law used by \citet{Schlafly11},

\begin{equation}
\frac{A_V}{E(B-V)}=R_V,
\end{equation} 
where $R_V=3.1$, the colour excess $E(B-V)$ is proportional to the $V$ band extinction caused by the Galactic dust. That is, from an $A_V$ value given in Table 1, a colour excess $E(B-V)$ for a system exists. With a known colour excess, $(B-V)_0$ colour of a binary would be known, according to the definition:

\begin{equation}
E(B-V)=(B-V)-(B-V)_0,
\end{equation}
where $(B-V)$ is the observed and $(B-V)_0$ is de-reddened colours of the system. Since Eq. (11) is valid for either one of the components of the binary, the same colour excess $E(B-V)$ could be used to calculate de-reddened colour of any component. Colour contributions of components are available only if the light ratio ($L_2/L_1$) in the $B$ band is available together with the light ratio ($L_2/L_1$) in the $V$ band. Therefore, computed de-reddened colours are given in Table 4 only for those systems if ($L_2/L_1$) is available for both $B$ and $V$ bands.

The de-reddened colours and effective temperatures of the stars in this study are plotted in Fig. 3. We could not find a simple continuous function, e.g. $\log T_{eff}=f[(B-V)_0]$, to fit the data for the full range of observed temperatures. This is because, it is clear on the figure that there is a break point at 10000 K ($\log T_{eff}=4$). Consequently, a linear function 

\begin{equation}
\log T_{eff}=-2.03406(0.13565)\times(B-V)_0+3.89169(0.02717),  
\end{equation}
was found best to express the data if $\log T_{eff}>4$, by the least squares method, with a standard deviation $\sigma=0.084$ and correlation coefficient $R^2=0.757$. For the stars with $\log T_{eff}\leq 4$, a quadratic function   

\begin{eqnarray}
\log T_{eff}=0.07569(0.012)\times(B-V)^2_0\nonumber \\
-0.38786(0.01368)\times(B-V)_0+3.96617(0.00338),  
\end{eqnarray}
was found best to express the data by the least squares method, with a standard deviation $\sigma=0.095$ and correlation coefficient $R^2= 0.941$. The numbers in the parenthesis in both functions are the internal random errors of the coefficients determined by standard error analysis techniques of the least square methods.

\begin{figure}
\begin{center}
\includegraphics[scale=0.80,angle=0]{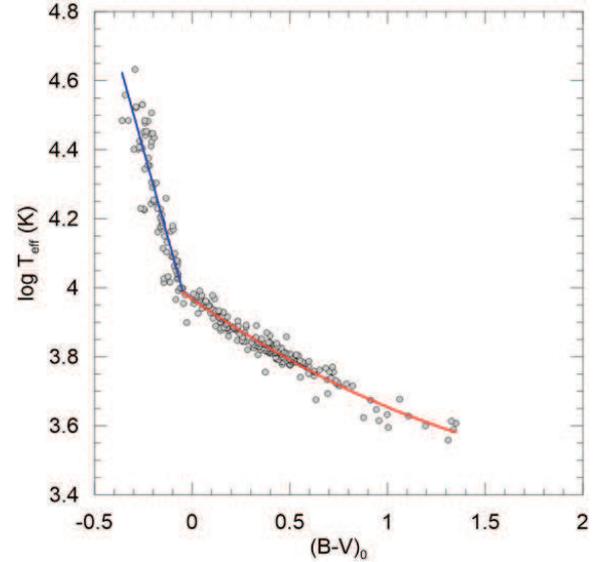}
\caption[] {De-reddened colours ($(B-V)_{0}$) and published effective temperatures ($\log T_{eff}$) of stars in this study if the light ratio ($L_2/L_1$) is available both $B$ and $V$ bands (circles). The best fitting linear function (if $\log T_{eff}> 4$) and the best fitting quadratic function (if $\log T_{eff}<4$) and the break point at $\log T_{eff}=4$ is clear.}
\end{center}
\end{figure}

\subsection{$BC-T_{eff}$ relation}
The $BC$ coefficients listed in Table 4 are shown together with the best fitting curve in Fig. 4. We have tried various degrees of polynomials to fit computed $BC$ data by the least squares method and found out that the best fitting function is a fourth degree polynomial with a standard deviation $\sigma=0.215$ and correlation coefficient $R^2=0.941$. The coefficients and associated uncertainties (Table 5) are determined by the least squares method and according to standard error analysis techniques of the least squares. 
 
\begin{figure}
\begin{center}
\includegraphics[scale=0.70,angle=0]{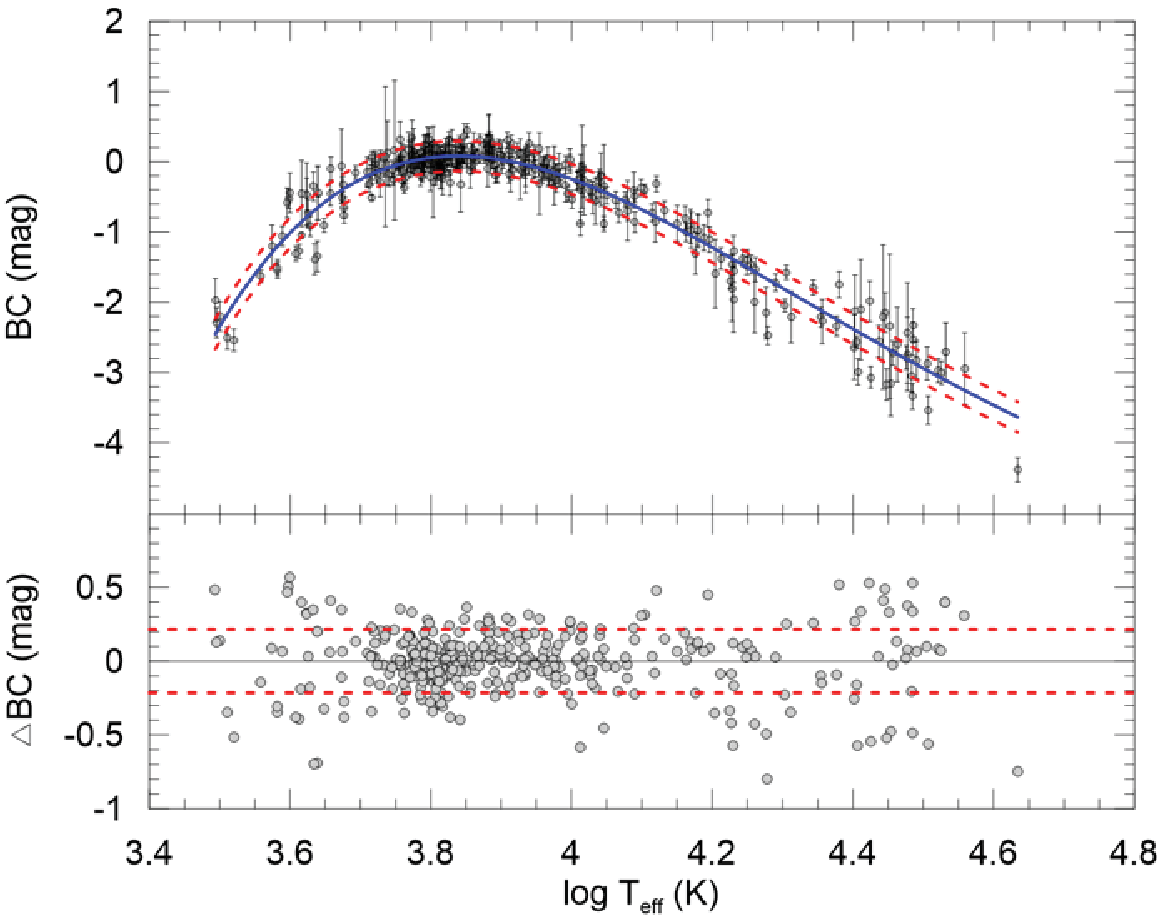}
\caption[] {Computed $BC$ coefficients of 400 stars and best fitting curve (solid). Dotted lines are one sigma confidence limits (upper) and one-sigma deviations of data points from the $BC-T_{eff}$ curve (lower).}
\end{center}
\end{figure}

\begin{table*}
  \caption{Coefficients and associated uncertainties of the $BC-T_{eff}$ relation, which is a 4th degree polynomial, representing the main-sequence stars in the Solar neighbourhood.}
    \begin{tabular}{ccccc}
\hline
\multicolumn{5}{c}{$BC=a+b\times(\log T_{eff})+c\times(\log T_{eff})^2+d\times(\log T_{eff})^3+e\times(\log T_{eff})^4$}\\
\hline
$a$ & $b$ & $c$ & $d$ & $e$ \\ 
\hline
    -2360.69565 & 2109.00655  & -701.96628  & 103.30304  & -5.68559 \\ 
    (519.80058) & (519.47090) & (194.29038) & (32.23187) & (2.00111)\\ 
\multicolumn{5}{c}{$\sigma = 0.215$,~~~~~~$R^{2} = 0.941$}  \\ 
\multicolumn{5}{c}{valid in the range $3100 < T_{eff}\leq 36000$ K}  \\
\hline
$BC=0.00$           & $T_{eff,1}=5859$ K   & $T_{eff,2}=8226$ K &  & \\
$BC_{max}=0.095$          &   $T_{eff}=6897$ K   &                    &  & \\
$BC_{\odot}=-0.016$ & $T_{\odot}=5772$ K   &                    &  & \\
\hline
    \end{tabular}%
\end{table*}%

According to Table 5, the $BC-T_{eff}$ relation found in this study has two roots, that is, it cuts $BC=0.00$ line twice at $T_{eff,1}=5859$ K, and $T_{eff,2}=8226$ K, thus the $BC$ values are positive ($BC>0$) between the temperatures $5859<T_{eff}<8226$ K. The maximum value $BC_{max}=0.095$, occurs at $T_{eff}=6897$ K. Bolometric correction for Solar-type stars with $T_{eff}=5772$ K is found to be $BC=-0.016$. 

Readers should keep in mind that the $BC$ values computed according to the $BC-T_{eff}$ relation presented in Table 5 are for the main-sequence stars with Solar metallicity, $0.008\leq Z\leq 0.040$ corresponding to iron abundances $-0.28\leq \rm{[Fe/H]}\leq0.61$ dex \citep{Eker18}. Spectral type and $\log g$ distributions related to the $BC$ values will be discussed next.

\subsection{Fundamental astrophysical parameters related to $BC$}

A typical mass and a typical radius of a main-sequence star of a given effective temperature, within the Galactic disc at the Solar neighborhood with metallicities in the range $0.008\leq Z\leq 0.040$, were computed by using the interrelated MLR, MRR and MTR functions by \citet{Eker18}. Because the same sample of stars are used in this study to establish relations between effective temperatures related to un-reddened $B-V$ colours (Eqs. 12, 13) and BCs, we have completed Table 6 for interested readers who are looking for a single table listing spectral types, effective temperatures, intrinsic $B-V$ colours, bolometric corrections, absolute $V$ magnitudes, masses, radii and surface gravities.

\begin{table}
\setlength{\tabcolsep}{5pt}
\scriptsize
  \caption{Fundamental astrophysical parameters related to $BC$ for nearby main-sequence stars with $0.008\leq Z\leq 0.040$.}
    \begin{tabular}{cccccccc}
\hline
    SPT & $T_{eff}$ & $(B-V)_0$ & $BC$ & $M_V$ & $M$ & $R$ & $\log g$ \\
      & (K) & (mag) & (mag) & (mag) & ($M_{\odot}$)  & ($R_{\odot}$) & (cgs) \\
\hline
    O2 & 52483 & -0.33* & -4.050 &-6.913 &63.980 &16.734 & 3.797 \\
    O3 & 46990 & -0.33* & -3.821 &-5.996 &44.260 &12.312 & 3.904 \\
    O4 & 43251 & -0.33* & -3.645 &-5.424 &34.809 &10.301 & 3.954 \\
    O5 & 40738 & -0.33* & -3.516 &-5.057 &29.669 & 9.236 & 3.980 \\
    O6 & 38282 & -0.33* & -3.379 &-4.708 &25.380 & 8.362 & 3.998 \\
    O7 & 35810 & -0.326 & -3.229 &-4.365 &21.660 & 7.616 & 4.011 \\
    O8 & 33963 & -0.314 & -3.108 &-4.113 &19.215 & 7.131 & 4.016 \\
    O9 & 32211 & -0.303 & -2.984 &-3.878 &17.123 & 6.721 & 4.017 \\
    B0 & 29512 & -0.284 & -2.777 &-3.520 &14.277 & 6.171 & 4.012 \\
    B1 & 25119 & -0.250 & -2.382 &-2.947 &10.459 & 5.454 & 3.984 \\
    B2 & 21135 & -0.213 & -1.946 &-2.430 & 7.699 & 4.967 & 3.933 \\
    B3 & 18408 & -0.184 & -1.595 &-1.705 & 6.123 & 3.989 & 4.024 \\
    B5 & 15136 & -0.142 & -1.108 &-0.872 & 4.516 & 3.214 & 4.079 \\
    B6 & 13964 & -0.125 & -0.918 &-0.545 & 4.007 & 2.974 & 4.094 \\
    B7 & 13032 & -0.110 & -0.761 &-0.268 & 3.625 & 2.797 & 4.104 \\
    B8 & 12023 & -0.093 & -0.588 & 0.053 & 3.234 & 2.617 & 4.113 \\
    B9 & 10666 & -0.067 & -0.356 & 0.535 & 2.743 & 2.394 & 4.119 \\
    A0 &  9886 & -0.051 & -0.228 & 0.848 & 2.478 & 2.274 & 4.119 \\
    A1 &  9419 & -0.020 & -0.156 & 0.904 & 2.325 & 2.362 & 4.058 \\
    A2 &  9078 &  0.021 & -0.107 & 1.080 & 2.216 & 2.292 & 4.064 \\
    A3 &  8750 &  0.063 & -0.062 & 1.259 & 2.113 & 2.226 & 4.068 \\
    A5 &  8222 &  0.136 &  0.000 & 1.569 & 1.952 & 2.123 & 4.075 \\
    A6 &  7980 &  0.171 &  0.025 & 1.723 & 1.879 & 2.077 & 4.078 \\
    A7 &  7745 &  0.207 &  0.045 & 1.879 & 1.810 & 2.033 & 4.080 \\
    A8 &  7534 &  0.241 &  0.060 & 2.026 & 1.749 & 1.994 & 4.082 \\
    F0 &  7161 &  0.305 &  0.077 & 2.302 & 1.643 & 1.928 & 4.084 \\
    F1 &  6966 &  0.340 &  0.081 & 2.457 & 1.588 & 1.893 & 4.085 \\
    F2 &  6792 &  0.373 &  0.080 & 2.602 & 1.540 & 1.863 & 4.085 \\
    F3 &  6637 &  0.403 &  0.077 & 2.735 & 1.498 & 1.838 & 4.085 \\
    F5 &  6397 &  0.453 &  0.064 & 3.225 & 1.354 & 1.588 & 4.168 \\
    F6 &  6310 &  0.472 &  0.058 & 3.404 & 1.305 & 1.508 & 4.197 \\
    F7 &  6223 &  0.491 &  0.050 & 3.581 & 1.259 & 1.434 & 4.225 \\
    F8 &  6152 &  0.507 &  0.042 & 3.727 & 1.222 & 1.377 & 4.248 \\
    G0 &  6026 &  0.536 &  0.026 & 3.986 & 1.161 & 1.283 & 4.286 \\
    G1 &  5957 &  0.552 &  0.016 & 4.129 & 1.128 & 1.236 & 4.307 \\
    G2 &  5888 &  0.569 &  0.005 & 4.270 & 1.098 & 1.191 & 4.327 \\
    G3 &  5848 &  0.579 & -0.002 & 4.354 & 1.080 & 1.165 & 4.339 \\
    G5 &  5741 &  0.606 & -0.022 & 4.585 & 1.031 & 1.097 & 4.371 \\
    G6 &  5689 &  0.619 & -0.033 & 4.669 & 1.019 & 1.081 & 4.379 \\
    G7 &  5649 &  0.630 & -0.042 & 4.732 & 1.011 & 1.069 & 4.385 \\
    G8 &  5559 &  0.654 & -0.064 & 4.881 & 0.990 & 1.041 & 4.399 \\
    K0 &  5248 &  0.742 & -0.158 & 5.421 & 0.922 & 0.951 & 4.447 \\
    K1 &  5070 &  0.797 & -0.227 & 5.753 & 0.884 & 0.903 & 4.474 \\
    K2 &  4898 &  0.855 & -0.306 & 6.092 & 0.848 & 0.858 & 4.500 \\
    K3 &  4732 &  0.914 & -0.395 & 6.439 & 0.813 & 0.817 & 4.525 \\
    K5 &  4345 &  1.069 & -0.661 & 7.327 & 0.736 & 0.727 & 4.582 \\
    M0 &  3802 &  1.353 & -1.225 & 9.113 & 0.558 & 0.541 & 4.719 \\
    M1 &  3648 &  1.455 & -1.438 & 9.643 & 0.524 & 0.508 & 4.745 \\
    M2 &  3499 &  1.569 & -1.674 &10.185 & 0.492 & 0.479 & 4.769 \\
    M3 &  3350 &  1.704 & -1.943 &10.773 & 0.462 & 0.452 & 4.793 \\
    M4 &  3148 &  1.946 & -2.368 &12.101 & 0.323 & 0.338 & 4.890 \\
    M5 &  2999 &  2.243 & -2.736 &13.055 & 0.249 & 0.284 & 4.928 \\
\hline
    \end{tabular}%
\\
(*) values taken from \citet{Sung13}.
\end{table}%
        
Typical main-sequence masses for intermediate-mass and massive stars ($M>1.5 M_{\odot}$) are computed from the effective temperatures given in column 2 using the mass-temperature relation (MTR). Once masses are known, then typical radii are computed from the luminosities estimated from the MLR function, according to Stefan-Boltzmann law. Since MTR and MLR functions are given for the stars with masses up to 31 $M_{\odot}$, $M$ and $R$ in the first three rows are extrapolated values. The surface gravity in the last column is then computed from the masses and radii listed in columns 6 and 7. Unfortunately, there is no single MTR function for the stars with the masses less than 1.5$M_{\odot}$. For those stars, MLR and MRR relations were used to estimate typical $M$ and $R$ for the main-sequence stars by checking up consistency between $L$, $R$ and $T_{eff}$ in the relation $L=4\pi R^2\sigma T_{eff}^4$, where $L$ and $R$ comes from MLR and MRR respectively, by trying different values of $M$. 

Typical luminosities and corresponding absolute bolometric magnitudes are not listed in Table 6. Instead, absolute $V$ magnitudes, which are computed from absolute bolometric magnitudes according to Eq. (1), are given. The $BC$ values are calculated from the $BC-T_{eff}$ relation (Table 5) using $T_{eff}$ given in column 2. Despite the $BC-T_{eff}$ relation is valid for $3100<T_{eff}<36000$ K, the trend of the fitting curve (Fig. 4) encouraged us to extrapolate it for the first five rows and the last row rather than leaving those rows empty. That is, a single $BC-T_{eff}$ relation is used to calculate $BC$ of the main-sequence stars in the full range from the spectral type O2 to M5. While Eq. (12) is used to calculate intrinsic $B-V$ colours for early type hotter stars ($\log T_{eff}>4$) and Eq. (13) is used to calculate intrinsic $B-V$ colours for late type cooler stars ($\log T_{eff}<4$). The trend of intrinsic $B-V$ colours, however, did not encouraged us to extrapolate the linear relation (Eq. 12) to hotter stars, thus, $B-V$ colours of five rows in Table 6 are taken from \citet{Sung13}.
 
The columns of Table 6 are self consistent because all other columns are produced from column 2 by interrelated MLR, MRR, MTR \citep{Eker18} and the effective temperature, colour $(B-V)$ and bolometric correction relations found in this study. Correlation between the spectral types and the effective temperatures is not in the scope of this study. However, it is customary to include spectral types in the first columns of such tables. Therefore, we have used spectral types and effective temperatures of \citet{Sung13}.   

\section{Discussions}
\subsection{Comparisons to other determinations}
The $BC-T_{eff}$ curve determined in this study is compared to other determinations in Fig. 5. It appears to have a reasonable agreement with \citet{Johnson66} and \citet{Sung13} at low temperatures ($\log T_{eff}<3.86$), where the $BC$ values from \citet{Flower96} are smaller (more negative) from the lowest temperatures up to $\log T_{eff}\sim 3.6$ and the rest is about the same as the others up to $\log T_{eff}\sim3.9$. The $BC$ values of this study appear to agree \citet{Flower96} and \citet{Sung13} better than the $BC$ values of \citet{Johnson66}, which appear relatively larger at high temperatures ($\log T_{eff}>3.86$). A noticeably good agreement between this study, \citet{Flower96} and \citet{Sung13} at temperatures $\log T_{eff}\sim 3.96$ and $\log T_{eff}\sim 4.4$ deviates to be underestimated (more negative) away from these temperature points. $BC$ values of this study appears as if a straight line inclined towards lower (more negative) $BC$ values if $\log T_{eff}>3.96$. In this region, $BC$ values from \citet{Johnson66} are higher (less negative). 

\begin{figure}
\begin{center}
\includegraphics[scale=0.60,angle=0]{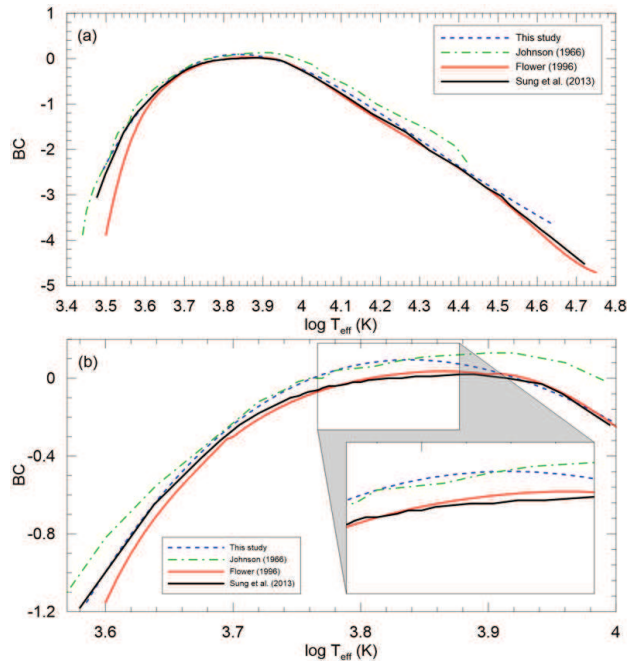}
\caption[] {The $BC-T_{eff}$ relation is compared to other determinations (a) and at $T_{eff,max}\approx 6897$ K where it is enlarged to show details (b). Smoothness of the curve of this study is clear while the others with a sudden jump at $\log T_{eff}=3.7$ \citep{Flower96}. Wavy appearance of \citet{Sung13} curve is caused by limiting $BC$ into two digits (enlarging window).}
\end{center}
\end{figure}

In order to see more details in the comparison, Fig. 4b shows an enlarged region near at $T_{eff, max}\approx 6897$ K. The smoothness of the curve of this study is clear, while the others show sudden jumps. The jumps of the curve by \citet{Flower96} is clear at the connecting points of the polynomials assigned by him in the temperature ranges ($\log T_{eff}\geq3.9$, $3.9<\log T_{eff}<3.7$, $\log T_{eff}\leq3.7$). The wavy appearance of the curve by \citet{Sung13} near the temperature maximum is caused by limiting the $BC$ values to two digits. To avoid such an effect, we preferred to adopt a $BC$ curve with $BC$ numbers which is in tree digits.

Fig. 6 shows the intrinsic $B-V$ colours of main sequence stars with metallicities $0.008\leq Z\leq 0.040$ in the Solar neighborhood which are compared to the intrinsic $B-V$ colours of the main-sequence stars according to \citet{Johnson66}, \citet{Flower96} and \citet{Sung13}. A reasonable agreement of the four curves is clear. 

\begin{figure}
\begin{center}
\includegraphics[scale=0.8,angle=0]{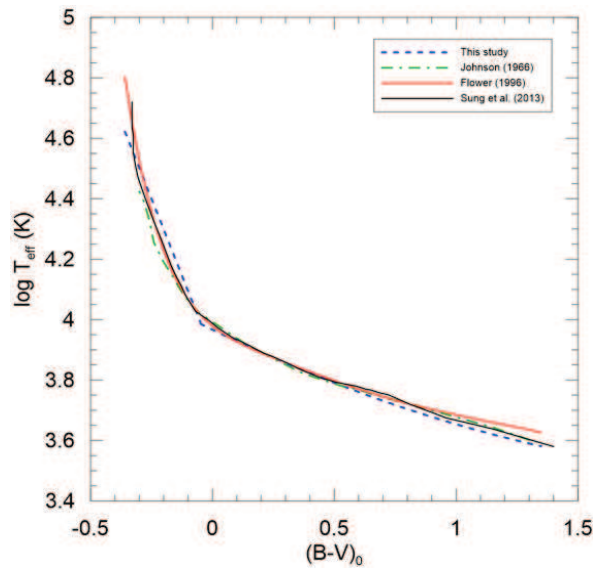}
\caption[] {Intrinsic $B-V$ colours of main-sequence stars in this study is compared to colours of \citet{Johnson66}, \citet{Flower96} and \citet{Sung13}.}
\end{center}
\end{figure}

\subsection{Importance of reddening corrections}
A reliable amount of reddening, in other words, an $E(B-V)$ colour excess of a star, is necessary for a reliable extinction. A reliable extinction is needed for a reliable absolute visual magnitude (Eq. 5). Being aware of the problem, all previous empirical $BC$ determinations \citep{Johnson64, Johnson66, Code76, Flower96} were done by assuming reddening is ignorable up to the distances 50-100 pc from the Sun \citep{Leroy93, Lallement19} because of the local bubble. ``Since most stars in these studies are bright and nearby, individual reddening values are generally very small and do not significantly contribute to the uncertainties in colours'' said \citet{Flower96}. Similarly, \citet{Code76} stated that ``$(B-V)_0$ is not an observed quantity for an individual star and uncertainties in interstellar extinction among our program stars contribute to the errors in determining a mean relation'', where $(B-V)_0$ is un-reddened colour of a star. 

We think, this study is the first, which is not ignoring interstellar reddening up to several kpc, when computing empirical $BC$ coefficients in order to calibrate a $BC-T_{eff}$ relation. The most distant binary considered here is one of the brightest early type systems with O type components V467 Vel which has a distance of 6720 pc according to its trigonometric parallax $\varpi=0.1488\pm0.096$ mas recorded in the {\it Gaia} DR2 \citep{Gaia18}. The biggest extinction $A_V$ used in this study belongs to an early type system MT91 372 with $A_V=7.01$ mag \citep{Kiminki15} which is a system having spectral types B1V+B5V at a distance of 1837 pc. 

Importance of reddening corrections is visualized in Fig. 7 where the predicted $BC-T_{eff}$ curve of this study is plotted together with the $BC$ values if interstellar extinction is ignored in a linear scale. The components of MT91 372 and MT91 696, which are the systems with highest extinctions, are indicated. Most of the systems with late type components ($T_{eff}<5000$ K) are nearby systems thus their extinction could be ignorable. However, for other systems, especially, for the systems with components having $T_{eff}>5000$ K, necessity of the extinction correction is clear. Notice, that almost all stars are located below the $BC-T_{eff}$ curve if $T_{eff}>15000$ K. 

\begin{figure}
\begin{center}
\includegraphics[scale=0.90,angle=0]{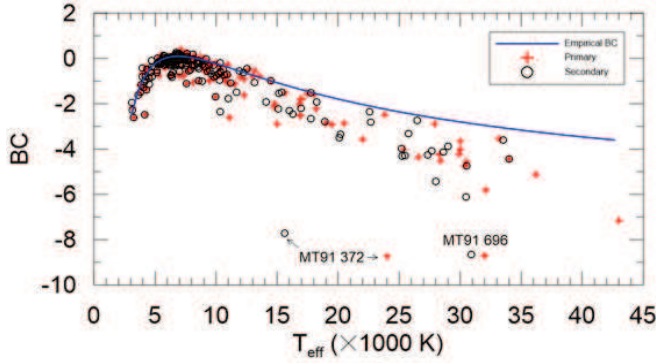}
\caption[] {$BC-T_{eff}$ relation (solid) is plotted together computed $BC$ values (dots) if interstellar extinction is ignored.}
\end{center}
\end{figure}

Distribution of $V$ band extinctions is displayed in Fig. 8. The eight systems with extinctions bigger than 2 mag are not shown. The 68.9\% of the whole sample have $V$ band extinctions smaller than 0.3 mag. Half of it, that is, 34\% of the systems have extinctions smaller than 0.1 mag. It is clear on Fig. 7 that if we also chose nearby stars to avoid extinction, $BC-T_{eff}$ relation would have been depending upon about 70 systems (34\% of the present sample) with components having effective temperatures less than 15000 K  

\begin{figure}
\begin{center}
\includegraphics[scale=0.90,angle=0]{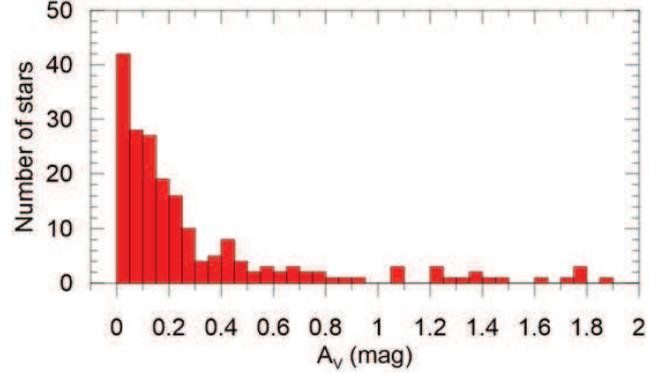}
\caption[] {The $V$ band extinction distribution of the present sample. The eight systems are not seen because their $A_V$ is bigger than 2 mag.}
\end{center}
\end{figure}

The extinctions ($A_V$) used in study are compared to the extinctions according to 3D reddening maps of \citet{Green19} in Fig. 9. Concentration of data along the diagonal indicates a reasonable correlation despite several systems are scattered more than an expectation. For the sake of consistency, we preferred not to use the 3D reddening maps of \citet{Green19} because of following reasons: 1) the method of computing $A_V$ is not applicable to all systems in this study. The 3D reddening maps of \citet{Green19} are available for $\delta >-30^{o}$. 2) The reddening maps, whether they are 2D or 3D, all gives a mean reddening towards a direction. However, there could be systems with circum-binary dust redder than the mean value. 

\begin{figure}
\begin{center}
\includegraphics[scale=.5,angle=0]{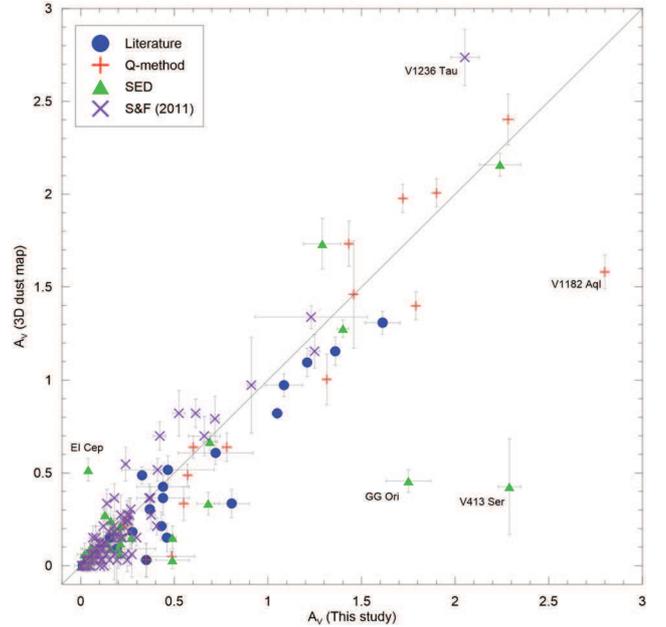}
\caption[] {Comparison of extinctions of this study and the extinctions according to the 3D reddening maps of \citet{Green19}.}
\end{center}
\end{figure}

It is not a coincidence nor due to large random observational errors that the most scattered systems from the diagonal on Fig. 9 are from the private determinations, $Q$ method or SED. Our $A_V$ values estimated from the 2D maps of \citet{Schlafly11} seems to have better correlation than the five most scattered data from private determinations.

\begin{figure*}
\begin{center}
\includegraphics[scale=1,angle=0]{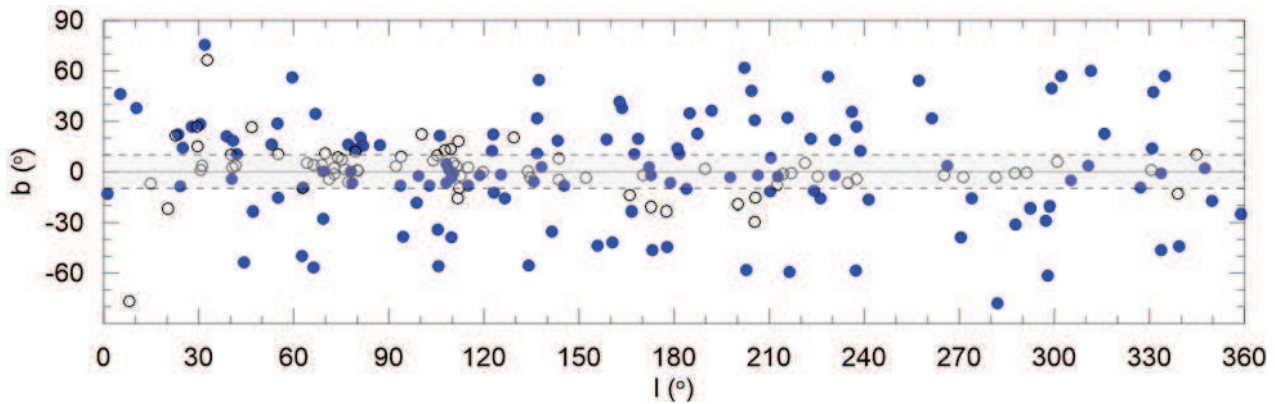}
\caption[] {Distribution of 206 systems on the Galactic coordinates. The filled symbols show the systems with extinction estimated according to Galactic dust maps \citep{Schlafly11} and the empty circles are the systems which have extinction by private determinations (see references in Table 1).}
\end{center}
\end{figure*}

Distribution of the present sample (206 systems) on the Galactic coordinates is given in Fig. 10, where the systems with extinction or $E(B-V)$ determined by various authors are shown by empty circles. The filled circles represents the systems $A_V$ calculated according to the method described NASA/IPAC Galactic Dust Reddening and Extinction maps of \citet{Schlafly11}. Notice that almost all or most of the empty circles are located on the Galactic plane or near to it ($|b|<5^{o}$). There are some systems shown by filled circles with $|b|<5^{o}$ on Fig. 10, which must be nearby systems in the local bubble with small and ignorable extinction \citep{Leroy93, Lallement19}. It is also possible that extinction of a system is caused by circumstellar dust and it is more than the dust predicted from 2D extinction map of the Galaxy. So that, there could be systems located at high Galactic latitudes with unusual reddening, so we preferred to use private determinations $E(B-V)$ rather than Galactic dust maps.

\section{Conclusions}

Here, $BC-T_{eff}$ and $(B-V)_0-T_{eff}$ relations for main-sequence stars in the Solar neighborhood having metallicities in the range $0.008\leq Z\leq 0.040$ equivalent to an iron abundance distribution $-0.28\leq [{\rm Fe/H}]\leq0.61$, are calibrated. The calibration sample is selected from detached eclipsing double-lined spectroscopic binaries because the most accurate mass, radii and temperatures could be obtained from their simultaneous LC and RV curves.

The $BC-T_{eff}$ and $(B-V)_0-T_{eff}$ relations were used together with interrelated MLR, MRR, and MTR relations from \citet{Eker18} and a self consistent table of photometric ($B-V$, $BC$, $M_{V}$) and physical ($M$, $R$, $\log g$) parameters of nearby main-sequence stars are prepared  (Table 6). It is a self consistent table because all the columns, except the first, are calculated analytically from typical $T_{eff}$ values of the  main-sequence stars.

Accuracy of $BC-T_{eff}$ relation would be improved greatly in the future together with the improvements of trigonometric parallax measurements, in accord with expectation of {\it Gaia} DR3, and upgrading the detached eclipsing double-lined spectroscopic data by increasing the quality and the number. 
 
\section{Acknowledgments}
Thanks to the anonymous referee whose comments were very useful. This work has been supported in part by the Scientific and Technological Research Council (T\"UB\.ITAK) by the grant number 114R072. Thanks to Akdeniz University BAP office for providing a partial support for this research. This research has made use of NASA's (National Aeronautics and Space Administration) Astrophysics Data System and the SIMBAD Astronomical Database, operated at CDS, Strasbourg, France and NASA/IPAC Infrared Science Archive, which is operated by the Jet Propulsion Laboratory, California Institute of Technology, under contract with the National Aeronautics and Space Administration. This work has made use of data from the European Space Agency (ESA) mission {\it Gaia} (\mbox{https://www.cosmos.esa.int/gaia}), processed by the {\it Gaia} Data Processing and Analysis Consortium (DPAC, \mbox{https://www.cosmos.esa.int/web/gaia/dpac/consortium}). Funding for the DPAC has been provided by national institutions, in particular the institutions participating in the {\it Gaia} Multilateral Agreement.

\end{document}